\documentclass[prb,twocolumn,longbibliography]{revtex4-1}
\pdfoutput=1
\usepackage{amsmath}
\usepackage{amssymb}
\usepackage{graphicx}
\usepackage{xcolor}
\usepackage{verbatim}
\usepackage{marginnote}
\usepackage{mathtools}
\usepackage{tabularx}

\newcommand\Gcal[1]{${\mathcal G}_{#1}$ }
\newcommand\Gcalnosp[1]{${\mathcal G}_{#1}$}

\begin{document}

\title{Hybrid grid/basis set discretizations of the Schr\"odinger equation}

\author{Steven R.\ White}
\affiliation{Department of Physics and Astronomy, University of California, Irvine, CA 92697-4575 USA}

\date{\today}

\begin{abstract}
We present a new kind of basis function for discretizing the Schr\"odinger equation in 
electronic structure calculations, called a gausslet, 
which has wavelet-like features but is composed of a sum of Gaussians. Gausslets are placed on
a grid and combine advantages of both grid and basis set approaches. They are orthogonal, infinitely
smooth, symmetric, polynomially complete, and with a high degree of locality.  Because they are formed
from Gaussians, they are easily combined with traditional atom-centered Gaussian bases. We also introduce
diagonal approximations which dramatically reduce the computational scaling of two-electron Coulomb
terms in the Hamiltonian.
\end{abstract}

\maketitle


\section{Introduction} 
Most numerical simulations of the Schr\"odinger cannot work directly in the continuum; they require a
discretization of some sort that maps the continuum to a finite number of degrees of freedom.
For example, in the many thousands of electronic structure calculations performed annually for 
molecules and solids, basis sets are usually used.  For molecules, these are usually atom centered 
functions composed of linear combinations of Gaussian type functions,
for which analytic integrals are used to quickly generate the
discrete Hamiltonian terms.  For solids, plane wave basis sets are
often used, which also have analytic integrals. These basis set methods have been developed and refined
over most of a century, and the software to use them has been improved to a remarkable degree.
Nevertheless,  a variety of alternative discretization methods, such as wavelet bases\cite{Harrison:2004,Yanai:2015,Mohr:2014} 
adapted grids\cite{Modine:1997,Beck:2000}, finite elements\cite{White:1989}, and sinc-function bases\cite{Jones:2016}
have continued to be developed. These alternatives have been pursued to address real weaknesses in existing
discretizations, but Gaussian and plane wave bases are still used for most calculations.

What is the motivation to try to make an improved discretization approach again? 
Our primary reason is that some of the most promising new strong-correlation
methods for electronic structure---the density matrix renormalization group
(DMRG)\cite{White:1992,White:1999,Chan:2011}, and related tensor network 
methods\cite{Nakatani:2013}---depend strongly on the
discretization. These methods have strong requirements for {\it locality} in
the basis, due to the low-entanglement nature of their approximations. 
One can use localized Gaussian basis sets for DMRG, but this combination does
not perform nearly as well as DMRG for simple lattice models, like the Hubbard
model.  The ideal basis for a DMRG or tensor network electronic structure
calculation would be some sort of hypothetical real space grid with a very
modest number of grid points, comparable
to the number of functions in an atom-centered Gaussian basis, and giving comparable accuracy. 
None of the current approaches approximate this ideal.  

Our approach has two key new elements.  The first is a novel type of basis function,
called a {\it gausslet},  related to the scaling functions of wavelet
approaches,  designed to live on a uniform grid. Several features of these functions
are particularly nice for electronic structure, particularly that they are
orthogonal, localized, smooth, symmetric, and composed of a sum of Gaussians. 
The second new element is a set of {\it diagonal }
 approximations for the Hamiltonian elements.  In making a diagonal approximation for the two electron
 interaction, the number of terms (integrals) needed drops from $N^4$ to $N^2$, where $N$ is the
 number of basis functions.  This reduction is not a large-$N$ asymptotic feature---it appears for
 any $N$ with no large coefficient in front of the $N^2$.  These two features take us a substantial way towards the performance of the hypothetical very coarse but accurate real space grid. 
 
The next section gives a more detailed overview of the problem and the approach. Section III discusses grids of Gaussians as bases, which have nearly ideal completeness properties but suffer from near linear dependence.  Section IV discusses ternary wavelet transformations which are used in
 Section V to derive gausslets.  
 Section VI presents diagonal approximations,  and Section VII has conclusions
 and a discussion of further directions. Throughout we give numerical examples in 1D.

\section{Overview}

Consider an arbitrary basis set approach.  The Hamiltonian is an operator with the coefficients of
the operator terms
defined by integrals over the basis.  By far
the most complex part of the Hamiltonian is the two-electron Coulomb interaction operator, parameterized by a four dimensional array or tensor
\begin{align}
    V_{ijkl} = \int_{\mathbf{r}_1}\int_{\mathbf{r}_2}
    \frac{\phi_i(\mathbf{r}_1)\phi_l(\mathbf{r}_1)
    \phi_j(\mathbf{r}_2)\phi_k(\mathbf{r}_2)}{|\mathbf{r}_1-\mathbf{r}_2|}
\end{align}
where the $\phi_i(\mathbf{r})$ make up the set of $N$ basis functions, $i=1\ldots N$. In some simpler
approximations (e.g. the local density approximation of density functional theory), 
one can bypass the use of $V_{ijkl}$ in favor of,
say, solving the Poisson equation, but usually for more accurate methods
treating electron-electron correlation explicitly, one cannot avoid
dealing with $V_{ijkl}$, which has $N^4$ elements.
 
In contrast, a simple cubic grid discretization gives a representation of the electron-electron interaction
which scales only as $N^2$, where now $N$ is the number of grid points. In such an approach,
one could use finite-difference approximations to represent kinetic energy derivatives.  
The nucleus-electron and 
electron-electron interactions would be evaluated point-wise, with
the two electron Hamiltonian terms taking the form $V_{ij} \hat n_i \hat n_j$, 
where $\hat n_i$ is the density operator on site $i$, and 
$V_{ij} = |\vec r_i - \vec r_j|^{-1}$ (with a suitable alteration at $i=j$).  

Simple grids like this are 
rarely used for electronic structure because the $N$ for an accurate 3D grid would be much 
bigger than for a typical basis.  Consider a single atom: the grid spacing needs to be set to resolve behavior near the nucleus, resulting in many grid points to describe the tails of the wavefunction 
far from the nucleus.  In contrast,   in a Gaussian basis the tails of the wavefunction can be described  using just a few basis functions.
For example, in an early  effort using a uniform 3D grid, utilizing finite elements rather than finite differences, as many as $10^5$ grid points were needed for accurate results for the molecule\cite{White:1989}.  Much more efficient approaches use adapted grids, putting more points near the nuclei\cite{Modine:1997}.  Neverthless, the number of grid points tends to be in the 1000's even for second row diatomic molecules. Wavelet bases are another approach with much of the locality of a grid,
and with adaptable increased resolution near nuclei, but again the number of functions tends to be
substantially larger than in Gaussian bases.   These types of methods can be useful for density functional theory or Hartree Fock calculations, but they are impractical for typical wavefunction methods treating correlation accurately.

Grid-like methods have a fundamental advantage for use with DMRG and other recently developed
tensor network methods:  these methods are based on the low entanglement of ground states when
expressed in a local real-space basis.  The
entanglement of ground states is governed by the \emph{area~law}
\cite{Eisert:2010,Hastings:2007}, which is specific to localized real space bases.
In a delocalized basis, a volume
law of entanglement holds instead, requiring the number of states $m$ kept to grow exponentially for fixed accuracy. 
Thus DMRG greatly prefers a real-space local representation. 
In addition, the calculation time for DMRG depends strongly on the number of two-electron interaction terms, 
so the $N^4$ scaling of $V_{ijkl}$ is a major drawback.  
The standard DMRG approach for molecules in a Gaussian basis 
utilizes a standard basis localization method before
DMRG is used.  This localization is less than ideal, and except in certain treatments of long chains, 
all $N^4$ two electron terms are kept.  

The very recently developed 
sliced basis DMRG approach (SBDMRG)\cite{Stoudenmire:2017} uses a finite-difference grid in one direction
(the long direction in a chain) and 2D Gaussians for the transverse directions. For chain systems, 
 the SBDMRG grid gives a reduction to $O(N^2)$ two electron terms, where $N$ scales linear with the length. SBDMRG also utilizes
a compression algorithm for the long range terms, yielding a  $O(N)$ calculation time, 
and chains of up to 1000
hydrogen atoms have been studied in the strongly correlated stretched-bond regime.  
While arbitrary molecules can, in principle, be treated with SBDMRG,
it is particularly suited for long chains, even more than standard QCDMRG.  One motivation for this work is
to find a good way to increase the grid spacing in SBDMRG without loss of accuracy, for faster and more efficient SBDMRG calculations. However, a more important motivation is to take  advantages of locality
in all three dimensions.

Here we present techniques which do  take us closer to an ideal combination of grid and basis discretizations.
Our approach is related to orthogonal wavelet bases.  
Wavelets have and continue to be used successfully in electronic
structure calculations, but our goals are different from many existing uses. 
A conventional wavelet basis can be
used in a way that has precise control over the accuracy, with systematically improvable accuracy, in, say,
a density functional theory calculation.\cite{Harrison:2004} 
These approaches make no use of standard Gaussian bases which 
give an excellent description of
core electrons with a very small number of functions.  Thus the price of the precise control of accuracy is a basis that may have  more than 
1000 functions for a small molecule,
which is not practical for most beyond-HF correlation calculations. 

Our interest is in these correlation calculations,
where one is interested in accuracies only up to about 1 mH, and for which Gaussian basis set descriptions of 
cores are fine, at least for energy differences. 
Thus we consider approaches where we use 
wavelet techniques in a very restricted way, combining them with
standard Gaussian basis sets.  A key consideration in developing such an approach is the calculation of integrals.
To make this easy, we construct functions similar to the scaling functions of wavelet transforms, but
out of an equal-spaced array of identical-width Gaussians.  
The resulting functions, called {\it gausslets}  have a number of significant advantages for electronic
structure calculations.

Gausslets live
on a uniform grid, with one function per grid point.  They are defined in 1D, but in 2 or 3D, one
simply takes products of the 1D functions. Gausslets are orthogonal and symmetric, and
almost compact in both real and momentum space, in the same sense that Gaussians are.
They have excellent completeness properties for representing smooth functions. All 
standard integrals have simple analytic forms, although one has to contract over the underlying
Gaussians.  A gausslet is shown in Fig. 1, together with seven of the Gaussians
of which it is composed. 
A 1D gausslet basis would be composed of this function plus integer
translations, and the whole basis can be scaled to a  grid spacing $a$.

\begin{figure}[t]
\includegraphics[width=6cm]{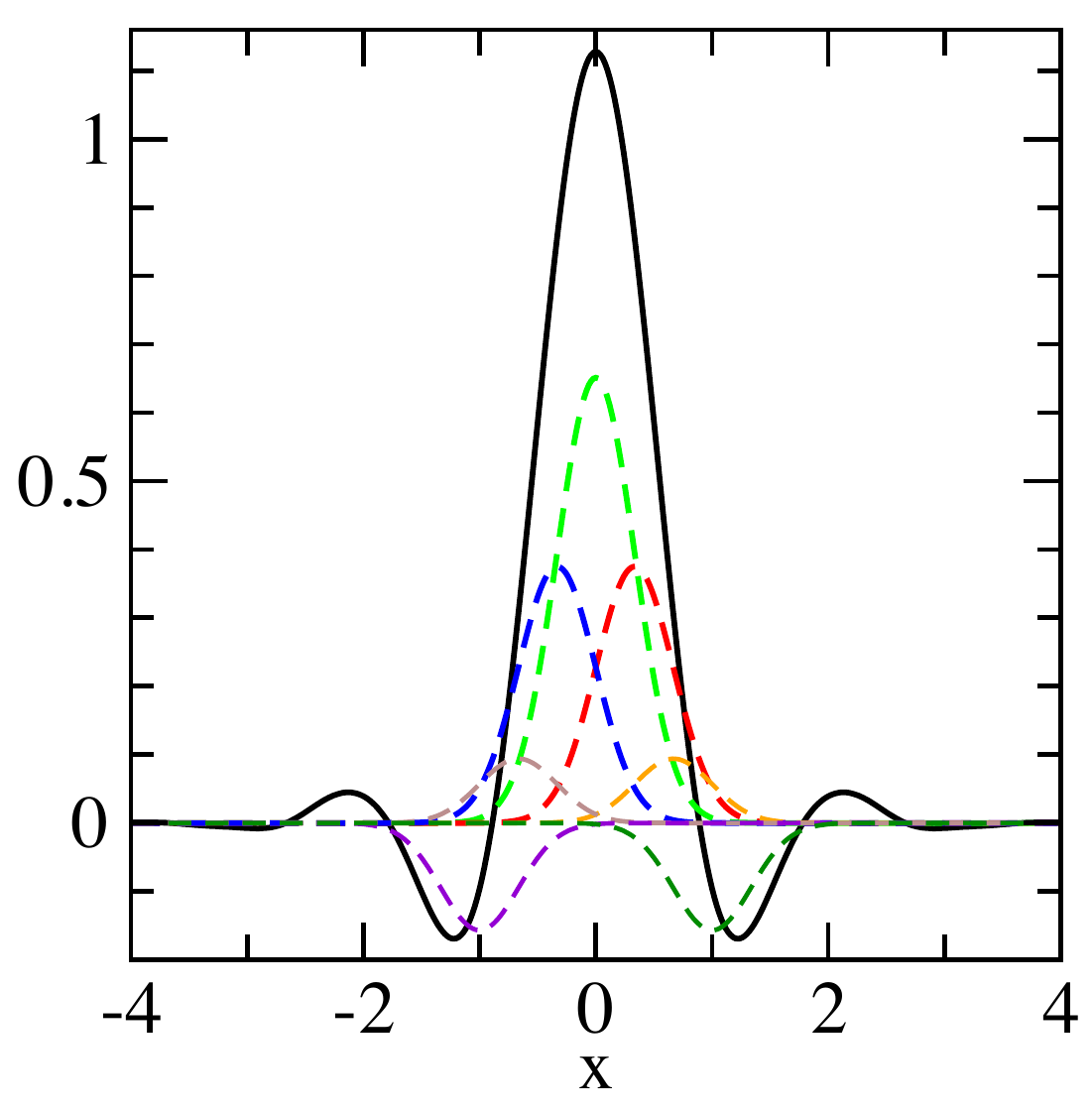}
\caption{The gausslet \Gcalnosp{6} (solid line), along with some of the Gaussians of which it is composed,
multiplied by their coefficients.
A basis is formed by translating  \Gcalnosp{6}  to center it on every integer, and then
scaling to any desired lattice spacing $a$ (here $a=1$). This basis is orthonormal
and can represent any polynomial up to fourth order. 
Gausslets  are exactly symmetric, and are defined as 
a sum of Gaussians of width
 $1/3$, with spacing $1/3$; \Gcalnosp{6} falls to about $10^{-12}$ at $x=\pm 12.5$. }
\label{fig:density}
\end{figure}

Gausslets are constructed using a recently developed class of wavelet transforms which
use a factor of three scaling transformation rather than the usual factor of two, in order
to make them exactly symmetric. They have an additional very important property: they integrate like a
$\delta$-function, for any low-order polynomial $p(x)$:
\begin{equation}
\int_{-\infty}^\infty dx \ {\mathcal G}(x-b) p(x) = p(b)
    \label{moments}
\end{equation}
This, in turn translates to reducing the two electron integrals from $N^4$ to $N^2$,  
$V_{ijkl} \to \tilde V_{ij}$, provided the interaction is smooth.  One way to make the interaction smooth would be
to replace the Coulomb electron electron interaction with a two-electron pseudopotential\cite{Prendergast:2001}.  

The first step in constructing gausslets is understanding the properties of
arrays of equally spaced, equal-width Gaussians.

\section{Arrays of Gaussians}
Arrays of Gaussians with identical widths are particularly convenient to use in constructing
basis functions because of their
analytic integrals, their smoothness, their completeness, and also because the product space
is greatly reduced and convenient.  The product of a Gaussian of width $w$ located at $i$, and
another of width $w$ at $j$, is a Gaussian centered at $(i+j)/2$ of width $w/\sqrt{2}$; the set
of all products of a grid of $N$ such Gaussians is a half-spaced grid with roughly $2N$ functions, 
rather than $N^2$ functions. The product space plays a central role in defining Hamiltonian matrix elements, so this simplification can significantly improve computational efficiency.

\begin{figure}[t]
\includegraphics[width=8cm]{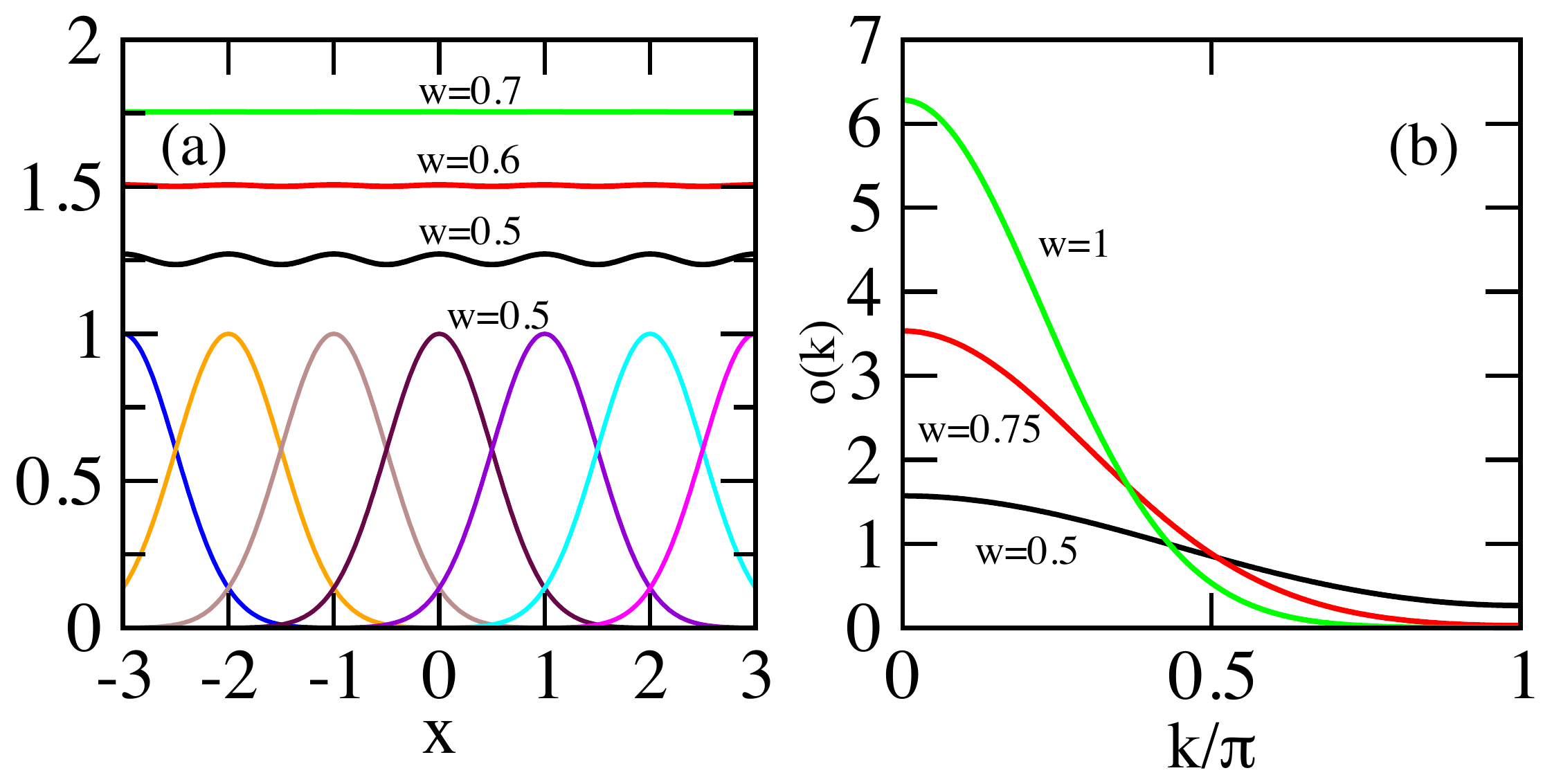}
\caption{(a) An array of Gaussians and their sums, $C(x)$.  (b) Overlap matrix eigenvalues.  }
\label{fig:gau}
\end{figure}

The completeness properties of arrays of Gaussians are interesting.
As is common in numerical analysis, we define completeness in terms of representing low order polynomials.
A grid of basis functions with good polynomial completeness is excellent at representing arbitrary smooth functions.
Consider a unit-spaced 1D grid,
and on each integer point $j$ put a Gaussian with width $w$
\begin{equation}
g_j(x) \equiv \exp[-\frac{1}{2}\frac{(x-j)^2}{w^2}]
\end{equation}
To see the completeness of the set $\{g_j\}$, consider the sum, shown in Fig. \ref{fig:gau}(a):
\begin{equation}
C(x) = \sum_j g_j(x).
\end{equation}
Since the $g_j$ are identical except for translation, the representation within this basis  of a constant
function must be proportional to $C(x)$. If $C(x)$ is not nearly constant, the basis is very poor---it can't even represent a constant.

We will show that $C(x)$ is very nearly constant for sufficiently large $w$. 
Note that $C$ is periodic, $C(x-1) = C(x)$, with period 1. Expand it in a Fourier series,
with coefficients $c(k=2\pi n)$,  where $n$ is an integer:
\begin{eqnarray}
c(k) &= &\int_0^1 dx\  C(x) \exp(-i k x)\\
&= & \sqrt{2 \pi} w  \exp(-\frac{1}{2} k^2 w^2). \nonumber
\end{eqnarray}
The largest deviation from constancy comes from the smallest nonzero $k$ in the series, namely $k = \pm 2\pi$, for which the factor 
$\exp(-\frac{1}{2}  k^2 w^2)$  is $\exp(-2 \pi^2 w^2)$. For
 $w = \sqrt{2}$, this factor is less than $10^{-17}$.  For $w=1$, it is less than $10^{-8}$.  
Thus for modest width $w$, $C(x)$
is exactly constant to double precision accuracy (15 or 16 digits).  In what follows,
we choose $w=1$, since this lack of completeness at the $10^{-8}$ level appears in the wavefunction.
According to standard variational arguments, it translates to errors in the energy of order $10^{-16}$. 
Note that higher terms in the series have much smaller coefficients, so that
not only is $C(x)$ nearly constant, a number of higher derivatives are nearly zero.  We will not try to
make this particularly quantitative, since we can test any such property with a trivial numerical
evaluation.

The flatness of $C(x)$---for the special case of Gaussians---also implies higher order polynomial completeness. 
In particular, we will show that the near constancy of $C(x)$ implies that, for $n \ge 1$, 
$\sum_j j^n g_j(x)$ is very nearly a polynomial in $x$ with order $n$.  (We have already shown it is
true for $n=0$.)  If {\it some}  linear combination of the $g_j$ can represent  an order-$m$ polynomial
for $0 \le m \le n$, then one can find a suitable linear combination that
can represent {\it any}  polynomial up to order $n$.
Gaussians have the property that their derivatives, to all orders, can be written as the same Gaussian times
a polynomial.  In particular, let
$P(x,n)$ denote a polynomial in $x$ of degree $n$; the nth derivative of $g_j(x)$  can be written as
\begin{eqnarray*}
g_j^{(n)}(x)   &=& P(x-j,n) g_j(x)  	\\		
&=& \sum_{l,m} C_{l,m} j^l x^m g_j(x)
\end{eqnarray*}
where $C_{l,m} =0$ if $l+m > n$.
The nth derivative of $C(x)$ is 
\begin{equation}
0 \approx C^{(n)}(x) = \sum_{l,m} C_{l,m} x^m \sum_j j^l g_j(x).
\end{equation}

In order to prove our result by induction, we pick out the $l=n$ term, for which $m = 0$,  and 
assume that for $l < n$, 
$\sum_j j^l g_j(x)$ is a polynomial $P_l(x)$. Then
\begin{equation}
C_{l,0} \sum_j j^n g_j(x) \approx -\sum_{l<n,m} C_{l,m} x^m P_l(x).
\end{equation}
Since $C_{l,m} =0$ if $l+m > n$, this means that if $\sum_j j^l g_j(x)$ is very nearly an order-$l$ polynomial for all $l < n$, then it is an order $n$ polynomial for $l=n$.  By induction, this is true
for any $n$.  This result breaks down when $C^{(n)}(x)$ stops being zero for large $n$.
For $w\gtrsim 1$ , the breakdown does not occur very quickly, and an array of Gaussians has
approximate polynomial completeness to high order.

The weakness of a grid of Gaussian as a basis is its lack of orthogonality, and that orthogonalizing
the functions involves a fairly singular matrix, associated with a near lack of linear independence. 
Let $S(j,k) = S(j-k) = \langle g_j | g_k \rangle$ be the overlap matrix of the $\{g_j\}$.  
We can form an orthonormal set of functions (all related by integer translations) 
using $q(j-k) \equiv S^{-1/2}(j,k)$ as
\begin{equation}
G_j(x)  \equiv \sum_k q(j-k) \ g_k(x).
\label{Gj}
\end{equation}
However, $S$ becomes increasing singular as $w$ increases.  The eigenvectors of $S$ are plane waves,
and the most singular point is at momentum $\pi$; this corresponds to near cancellation of $g_j$.  For
example, for $w=\sqrt{2}$, the unnormalized fit to a momentum $\pi$ plane wave,  $\sum_j (-1)^j g_j(x)$, is 
only of order $10^{-4}$ in magnitude.  The near singularity of $S$ means that the $G_j$ have long tails and widely
varying coefficients for the component Gaussians, making them computationally poorly behaved.

Fortunately the orthogonalization does not spoil the completeness, nor would a generic  convolution of the form Eq. (\ref{Gj}), transforming $g \to G$ using an arbitrary vector $q$.   
Consider
\begin{equation}
    \sum_j P(j,n) G_j(x)  = \sum_k \left[ \sum_j q(j-k) P(j,n)\right] g_k(x).
\label{Gjconv}
\end{equation}
The term in square brackets is a polynomial in $k$ of degree $n$, so the entire right side is a polynomial
in $x$ of degree $n$. By the same reasoning as above, this implies that the $G_j$ have the same excellent
completeness properties as the $g_j$.  

In summary, a grid of Gaussians has excellent completeness, as well as several other very desirable
features,  but poor orthogonality and linear independence properties.  Fortunately, as we show in the next two sections, these faults can be
fixed using wavelet technology.

\section{Ternary Wavelet Transformations}

In conventional compact orthogonal wavelet theory, as pioneered largely by Daubechies,\cite{Daubechies:1992} it is not possible
to have a symmetric, compact, orthogonal wavelet transformation (WT).  However, with ternary WTs , 
which change scales by a factor of 3 instead of 2, symmetry, orthogonality, and compactness
are compatible. Recently Evenbly and White (E\&W)\cite{Evenbly:2016} introduced new families of ternary WTs
with excellent completeness, compactness, and smoothness properties, which we will
make use of and extend.  A very carefully chosen set of coefficients
$c_k$,  $-m \le k \le m$, with $c_{-k} = c_k$, defines a symmetric wavelet transform.  Given a single
function $f(x)$, we define all integer translations  $f_j(x) = f(x-j)$ to form a basis set that lives on
an integer grid. Then the
wavelet transform produces a new basis set, also living on a grid with unit spacing, defined by a  function $f'$, defined by
\begin{equation}
f'(x) = \sum_k c_k f(3x-k) 
\label{ckf}
\end{equation}
The scaling function of the WT is the fixed point of Eq. (\ref{ckf}).  The fixed point exists
provided the initial $f$ satisfies some simple conditions, most importantly that they sum
to a constant (exactly),  $\sum_j f_j(x) = {\rm const}$.  Thus any functions defined by
Gaussians with finite $w$ are technically not suitable for generating the fixed point.  However, if
one only wants to apply the WT a modest number of times, Gaussian derived functions
can be an excellent starting point. Since we do not plan to use the wavelets to represent sharp core
functions (instead using standard Gaussians), a full multiscale resolution analysis is not needed, and for our uses, the fixed point is entirely unnecessary.

E\&W showed that WTs have a direct correspondence with quantum circuit
theory.  The circuit corresponding to a WT is defined by a small number of angles $\theta_k$;
each angle defines a unitary ``gate'', which is a $2\times2$ or $3\times3$ unitary matrix.  
The unitarity of the circuit is independent of the values of $\theta_k$, removing an annoying
set of constraints when optimizing a WT.  The circuits express symmetry much more naturally
than in conventional approaches.  Some of the symmetric ternary WTs constructed by E\&W appear to have better properties than any previous such WTs. 

Besides the transformation of the scaling function in Eq. (\ref{ckf}), the WT produces two
wavelet functions per scaling function. The scaling function captures the lowest momenum
behaviour, while the wavelets describe high momentum. 
These wavelets are less central to our discussion here.  

In the terminology of E\&W,
we consider here only Type I site-centered symmetric ternary wavelets.  These are characterized
by the number of low and high frequency moments associated with the scaling functions.  We
make this more specific here with the notation $W_{nlh}$ to describe the WT with $n$ angles,
$l$ low moments, and $h$ high moments.  The number $n$ also gives the number of layers of the circuit,
fixing the range of the $c_k$, specifically from $-(3n-2)$ to $3n-2$, with $c_{-k} = c_k$.  
The number $l$ gives the completeness of the scaling functions; more specifically, 
$l-1$ is the maximum polynomial order the scaling functions fit exactly, which is imposed
by making the wavelets orthogonal to any $l$th degree polynomial.
Similarly, $h$ controls the smoothness; specifically, $h-1$ 
gives the number of sign-flipped polynomials $(-1)^k k^h$ the scaling functions
are orthogonal too, which also corresponds to the Fourier transform of the $c_k$ having vanishing
value and $h-1$ derivatives at maximum momentum $\pi$. 

A related and important property of the corresponding scaling functions is that 
they integrate polynomials like a $\delta$-function.  This is associated with a property of the $c_k$:
\begin{equation}
    \sum_k c_k k^m = \sqrt{3} \delta_{m,0}
\label{ckm}
\end{equation}
for $m=0,\ldots,p-1$ for some $p$.  (The evenness of the $c_k$ makes this statement trivial for odd $m$.) 
The $\delta$-function order $p$ is related to the other moments, but not in a simple way:  we have found WTs
with the same $n$, $l$, and $h$, but with different $p$.  Usually, however, a large $p$ appears ``for free'' from
optimizing $l$ and $h$.

E\&W give just a few examples of these types of WTs, and not to full double precision.
Determination of a $W_{nlh}$ involves a nontrivial nonlinear optimization; we have found 
$W_{nlh}$ to high precision both for the examples of E\&W and a number of additional WTs with
higher order. All useful WTs had even $n$, and it is difficult  to converge useful WT for higher $n$.
Angles for the most useful WTs are listed in Table I.  These emphasize completeness over smoothness, 
but not completely, mostly setting $h=2$ (which is only one constraint, since the order-1 high moment is
automatically satisfied because of symmetry). The scaling functions with $h=2$ are nicely smooth; for $h=0$ they
are much more irregular.
This leaves $n-1$ degrees of freedom for completeness,
i.e. $l=n-1$.  Below we construct gausslets out of $W_{432}$, $W_{652}$,  $W_{872}$, and $W_{1092}$.
These WTs all have impressively
high $p$'s given by a simple formula: $p=2l$. 

\begin{table}[!htb]
    \begin{tabular}{|c||r|r|}
	\hline
	& \phantom{xxx} $W_{212}$ \phantom{xxx}&\phantom{xxx} $W_{220}$ \phantom{xxx}\\ \hline
	$\theta_1 $ & 0.16991845472706096855    & 0.27564279921626540397 \\     
	$\theta_2 $ & 0.78539816339744830962    & 0.67967381890824387419 \\
	\hline
    \end{tabular}

    \begin{tabular}{|c||r|r|}
	\hline
	& \phantom{xxx} $W_{432}$ \phantom{xxx}&\phantom{xxx} $W_{652}$ \phantom{xxx}\\ \hline
	        
	$\theta_1 $ &  0.33591409249043635638 & 0.47808035535078662712 \\
	$\theta_2 $ & -1.46977679545346969482 & 0.61724603408318791423 \\
	$\theta_3 $ & -0.16599563776337538784 &-1.39526939914470111011 \\
	$\theta_4 $ &  2.25517495885091800443 &-0.51453080478199670477 \\
	$\theta_5 $ & - 		      & 1.08710749852097545154 \\
	$\theta_6 $ & - 		      & 0.68268293409625710015 \\
	\hline
    \end{tabular}

    \begin{tabular}{|c||r|r|}
	\hline
	& \phantom{xxx} $W_{872}$ \phantom{xxx}&\phantom{xxx} $W_{1092}$ \phantom{xxx}\\ \hline
	$\theta_1 $ &  0.57548632554189299396 &   0.61183864100711914517 \\
	$\theta_2 $ &  1.07092896697683457430 &   0.70680189527339659367 \\
	$\theta_3 $ & -0.53397048757827478700 &  -2.76090222438702891461 \\
	$\theta_4 $ & -0.84404057490329784656 &   0.82725175790860767489 \\
	$\theta_5 $ & -2.19779794769420972567 &  -0.63493786130732787550 \\
	$\theta_6 $ & -0.20902293951451481799 &  -0.23346217196976042571 \\
	$\theta_7 $ &  2.32620056445765248695 &   1.21327891348912294041 \\
	$\theta_8 $ &  0.76753271083842639987 &  -1.15390181597860903352 \\
	$\theta_9 $ &  -		      &   1.74064098592517567308 \\
	$\theta_{10} $ & -		      &   0.63870849816381350029 \\
	\hline
    \end{tabular}
    \caption{Angles $\theta_k$ parameterizing  wavelet transforms of various depths.
    }
    \label{table:MaxTernary}
\end{table}

A WT can be applied to an array of Gaussians, producing a new basis where the functions are formed
from sums of Gaussians. 
The completeness properties of the Gaussians are transfered by the WT to the new basis, 
up to the completeness
order $l$.  To see this, let $i$ index a grid with spacing 1, and $j$ (an integer) run over a grid with spacing $1/3$.
Let $f_i(j) = c_{j-3i}$; we can think of $f_i$ as being a discrete basis function representing functions
living on the $1/3$ grid.  (Usually in wavelet theory one thinks of an infinite sequence of ever smaller
grids, which represents the continuum in the limit of infinitesimal grid spacing.  In contrast, we apply
the WT to a grid of functions which are already continuous functions of $x$.) The WTs have a 
discrete completeness property that means we can write
\begin{equation}
    j^{m} = \sum_i a_i f_i(j)
\end{equation}
for $m < l$.
Multiplying by $g_j(x)$ on both sides, and summing over $j$, we find that linear combinations
\begin{equation}
s_i(x) \equiv \sum_j f_i(j) g_j(x)
\end{equation}
can represent arbitrary low order polynomials.  Again, this implies that suitable linear combinations
of the $s_i(x)$ can represent any low order polynomial.
However, they are not orthogonal, which we address in the next section. 

\section{Gausslets}
We wish to use these properties of WTs and Gaussian grids to define convenient, orthogonal, complete functions
defined as sums of Gaussians. There are several possible ways to proceed. One way starts with the $s_i(x)$.
The $s_i(x)$ are not orthogonal, but their overlap matrix $S_s$ is much less singular than that
of the $g_j(x)$. The $s_i(x)$ look like somewhat broadened wavelet scaling functions, with oscillations that
partially cancel off-diagonal elements of the overlap matrix.  (Further applications of a WT would take them
closer to orthogonality.)  One can use $S_s^{-1/2}$ to orthogonalize the $s_i$.  The resulting functions
are close to what we want, but $S_s^{-1/2}$ introduces moderately long tails in the final functions. 
One can then apply a WT to these orthogonal functions, which would substantially decrease the tails, and
use those functions.  The main drawback is that the underlying Gaussian grid
would have a spacing of $1/9$ that of the final basis functions.  The more
Gaussians making up a basis function, the more computatonal 
work to use them, so we have developed a way to maintain the minimal $1/3$ spacing of the underlying Gaussians without the long tails.

\begin{figure}[t]
\includegraphics[width=6cm]{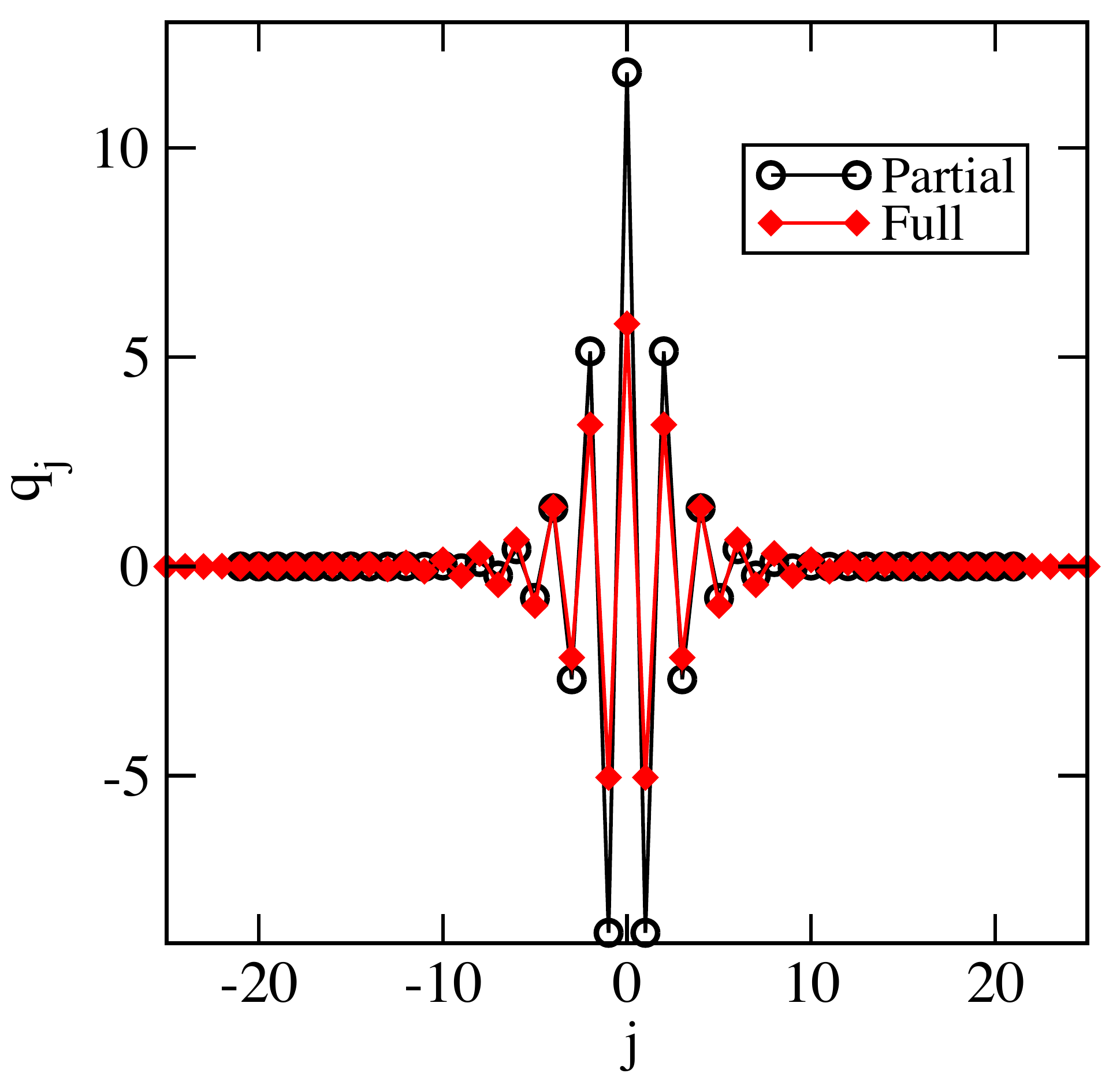}
\caption{Full and partial orthogonalizing vectors $q(k)$, for the case of \Gcalnosp{6}.  
The partial orthogonalizer extends out to $\pm 21$; the full orthogonalizer has fallen to about $10^{-12}$ at  $j=110$. }
\label{fig:qdecay}
\end{figure}

It is possible to partially orthogonalize 
the $g_j(x)$, so that the application of a particular WT  to the partially orthogonalized functions produces 
fully orthonormal functions.   Let function $G_j(x)$ be defined by Eq. (\ref{Gj}), for some vector $q(k)$, and then apply a WT (defined by $f_i(j)$):
\begin{equation}
    {\mathcal G}_i(x)  \equiv \sum_j f_i(j) \ G_j(x).
\label{Si}
\end{equation}
We know that a $q(k)$ exists that makes the ${\mathcal G}_i(x)$ orthonormal---we can use the overlap matrix $S_s$ 
to make $q(k)$, which completely orthonormalizes the $G_j(x)$, and thus the ${\mathcal G}_i(x)$. 
In that case, $q(k)$ falls to about $10^{-12}$ near $k=110$, so it is rather nonlocal, making
${\mathcal G}_i$ have long tails.
It is possible to find a much more compact $q(k)$ if it is optimized for a particular WT, requiring orthogonality
only of the ${\mathcal G}_i$ and not both the $G_j$ and the ${\mathcal G}_i$.
The $q$ convolution and the WT each take care of parts of the orthogonalization. A typical $q$ is
show in Fig. \ref{fig:qdecay}.
Finding such a $q$ is a rather tedious nonlinear optimization
problem, with many local minima. We implemented this optimization using a Nelder Mead algorithm, 
using restarts and  
high precision to deal with local minima, with the minimizations sometimes taking several days, and only achieving approximate double-precision orthogonality.  
A final orthogonalization 
step was performed to make the functions orthogonal to very high precision, with minimal impact on locality,
 using the inverse square root of the overlap matrix.
Fortunately, using these functions does not require that these optimizations be repeated; all one needs
are the final coefficients of the Gaussians, and the properties of the functions are easily verified.

This approach produces functions---gausslets---with a ratio of 3 between the spacing of the functions and
the underlying Gaussians. 
Normally with WTs there is a tradeoff between order and compactness.  The presence of the 
underlying Gaussians mean that
there is an inherent limit to the size
of a ``$1/3$-gausslet" which is greater than the width of the fixed point scaling function of the WT used
to make the gausslet.  However, the gausslets are substantially smoother.
We present coefficients for gausslets \Gcalnosp{4}, \Gcalnosp{6},
\Gcalnosp{8}, and \Gcalnosp{10}, where the index signifies $n$ of the underlying WT, $W_n$, which has $l=n-1$ and $h=2$.
The gausslet \Gcalnosp{6} is shown in Fig. 1; the others look similar, with slightly increasing width with the order.
The coefficients $b_j$ of these gausslets in terms of the underlying 
Gaussians are given in the Appendix in Table II-V.

\begin{figure}[tb]
\includegraphics[width=6cm]{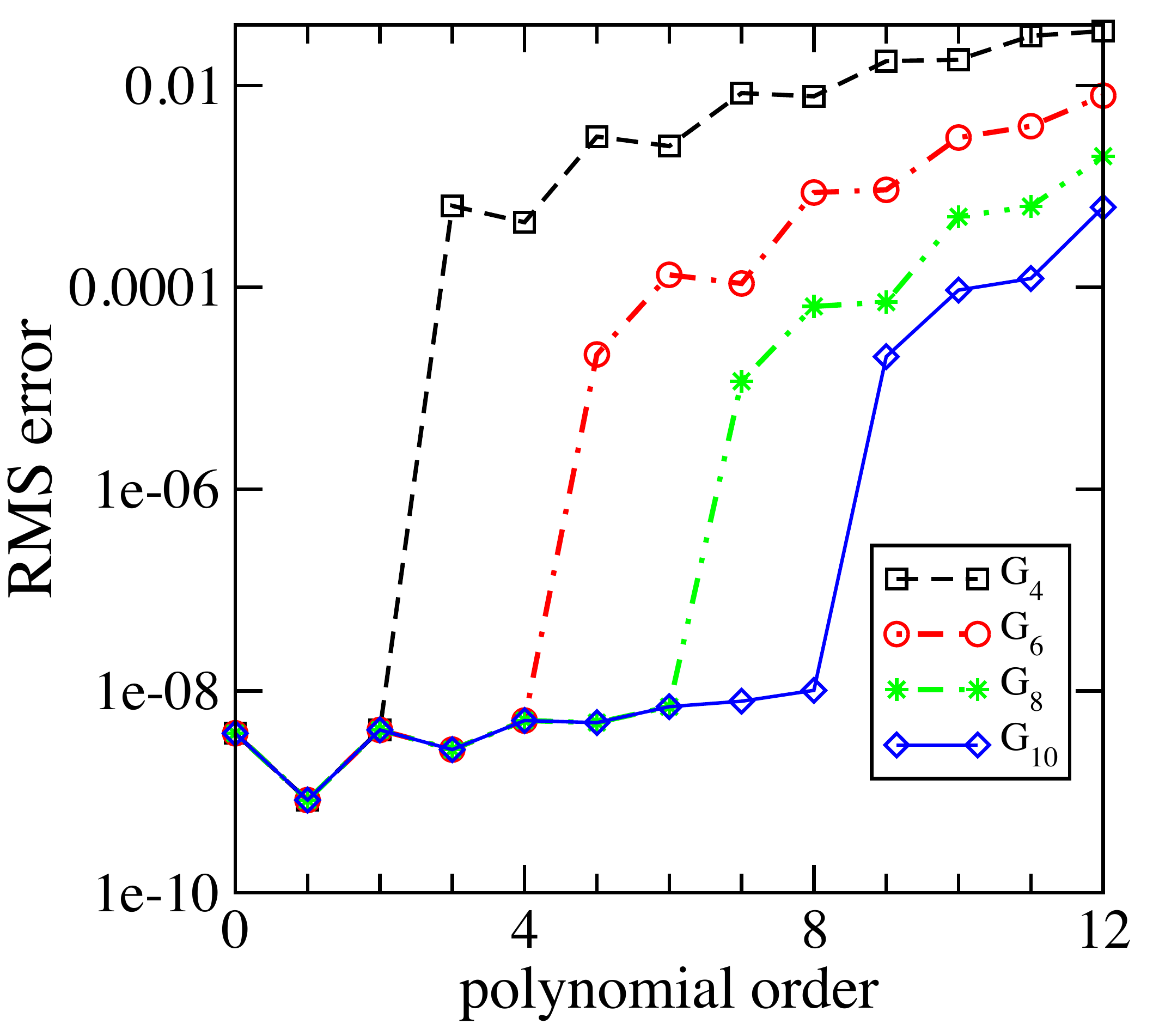}
\caption{Root mean square errors in the fit of an array of gausslets with unit spacing 
to scaled Chebyshev polynomials $T_n(x/10)$, over the range $0 \le x \le 1$,
for gausslets of different orders, as a function of $n$.   The gausslets of order $n$ are based on
wavelet transformations designed to
fit  polynomials of order   $0$ to $n-2$.  The residual error of order $10^{-8}$ visible in
the figure for small $n$ is due to the
finite width and imperfect completeness  of the underlying Gaussians making up the gausslets. Since
the underlying Gaussians are the same for all the gausslets, the residual errors are approximately the same.
The fitting errors of order $10^{-8}$ are expected to translate to energy errors of order $10^{-16}$.
 }
\label{fig:completeness}
\end{figure}

A simple test of the completeness of the gausslets is shown in Fig. \ref{fig:completeness}.  The gausslets are fitted to polynomials of various orders and the root mean square errors over the unit interval are shown. (The gausslets involved in the fitting extend well outside this interval, because of the small but finite tails of the gausslets.) The fitting coefficients were approximated
using the $\delta$-function property, as $T_n(i/10)$ for the gausslet centered at $i$, so this test
of the fitting also tests the $\delta$-function property.  Since the order of the $\delta$-function property is much
higher than for fitting property, one sees only the errors associated with the fitting.

\begin{figure}[t]
\centering
\includegraphics[width=8.3cm]{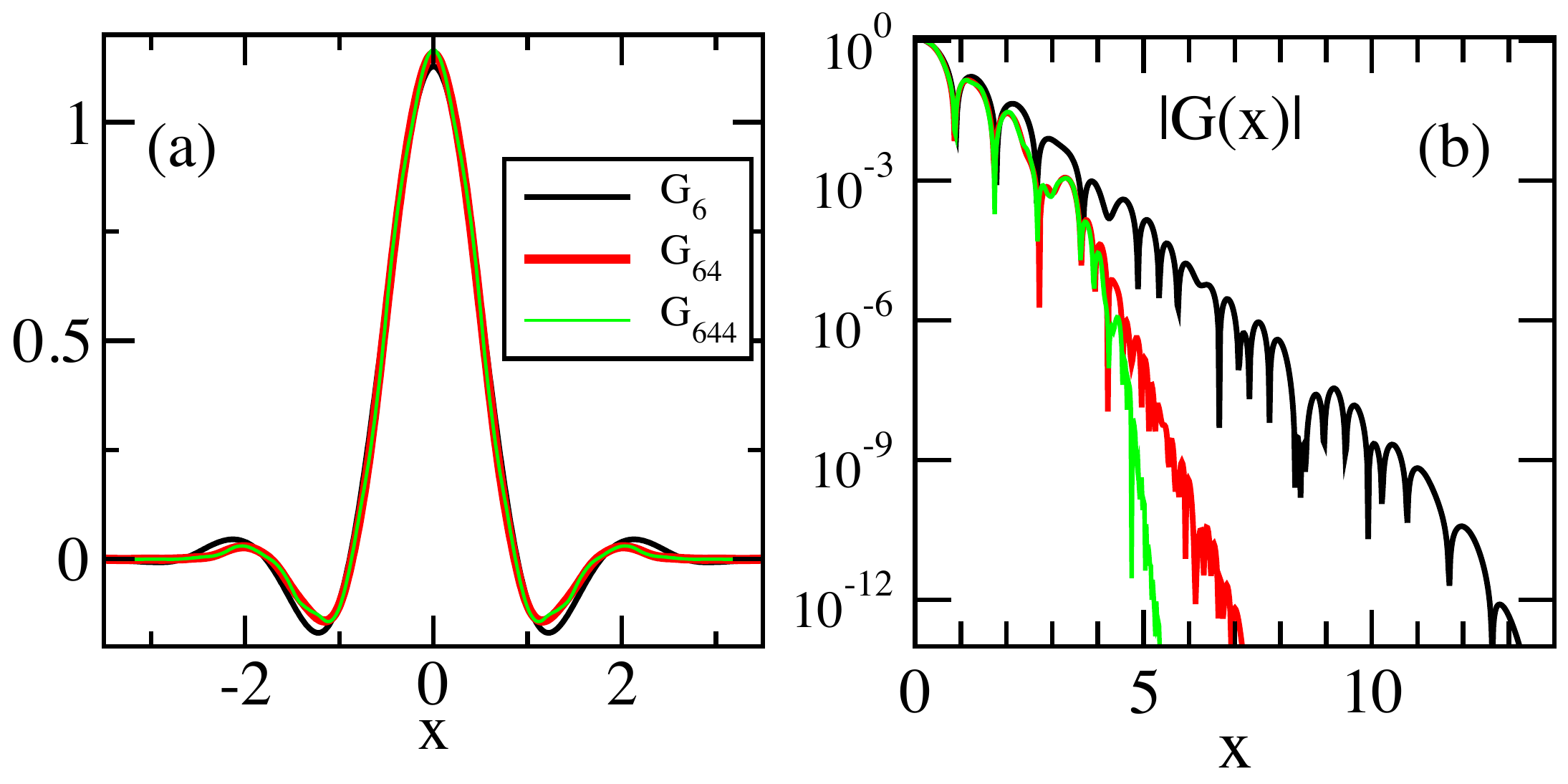} 
\caption{Comparison of a sequence of guasslets \Gcalnosp{6},  \Gcalnosp{6}$_4$,   \Gcalnosp{6}$_{44}$.  Repeated application of a wavelet transform shrinks the tails of the functions.}
\label{fig:Gdecay}
\end{figure}

One may wish to have more compact gausslets, paying the price of an underlying finer array of Gaussians.  
Suitable $1/9$, $1/27$, etc. gausslets are easily derived by applying WTs to one of the
\Gcalnosp{n}. We label these gausslets as in the following example:  \Gcalnosp{864} is formed from \Gcalnosp{8}
and applying $W_6$ and then $W_4$. These constructions also generate useful wavelet-like functions, and a 
multiresolution basis
can be formed for a modest range of scales, with each length scale having different functions.
If one successively applies $W_n$, the range of the functions decreases,
and the basis functions become close to the scaling function of $W_n$.
Alternatively, one can successively decrease the order over several steps, which produces very 
localized functions functions which are still smooth with a finite size underlying Gaussian array.
An example of this is shown in Fig. \ref{fig:Gdecay}.
A multiscale basis can be useful, but in many cases a better
alternative that produces smaller bases is to add atom-centered Gaussian functions to the gausslet
basis. The extra functions can be orthogonalized to the gausslets and themselves. A simple 1D example of
this approach is given below.

\begin{figure}[t]
\includegraphics[width=6cm]{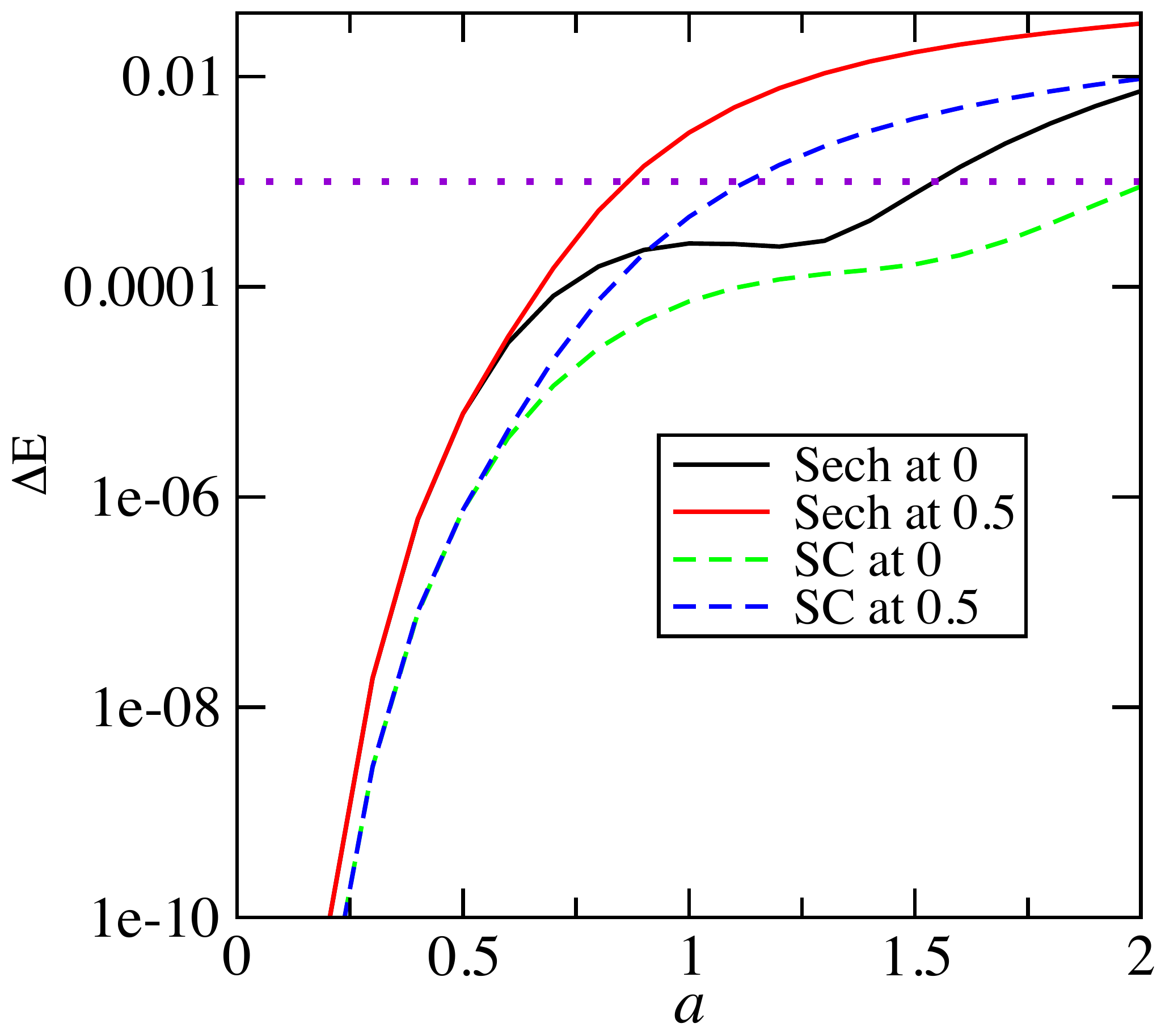}
\caption{Energy errors in solving 1D potentials with a \Gcal{10} gausslet basis with spacing $a$.
Here the ``Sech'' potential is the exactly solvable P\"{o}schl-Teller potential $V(x) = -{\rm sech}^2(x-b)$, for $b=0, 0.5$, and the ``SC'' potential is the soft Coulomb potential
$V(x)=-1/\sqrt{(x-b)^2+1}$, for $b=0, 0.5$.  The dotted line shows $\Delta E=10^{-3}$, a typical target accuracy for ``chemically accurate'' electronic structure calculations.
  }
\label{fig:figA}
\end{figure}

As a first test of the use of gausslets to solve the Shr\"odinger equation, we apply them to two different simple 1D smooth potentials; see Fig. \ref{fig:figA}.  The first, the P\"{o}schl-Teller potential, is exactly solvable, with energy $E=-1/2$.  The second, the
soft Coulomb potential, has been used as a 1D model with properties such as correlation energies roughly 
similar to that of real 3D molecules\cite{Stoudenmire:2012,Wagner:2012}. 
Here the SC potential would be that of pseudo-hydrogen. Each potential is centered at two positions to  demonstrate that the grid points need not be aligned with the potential centers.
To define the Hamiltonian matrices, we first evaluate the Hamiltonian
matrix elements for the underlying Gaussian grid with spacing $a/3$, analytically for the kinetic energy,
and numerically for the potentials. Then we transform to the gausslet basis using the $b_j$.
Both of these potentials have widths of order 1, and we see that very high accuracy is easily achievable for $a \sim 0.2$.  More important for our goals
is that one achieves accuracies of $O(10^{-3})$ for $a \sim 1$.  Large $a$ values are essential for applications to 3D. 

\begin{figure}[t]
\centering
\includegraphics[width=3.9cm]{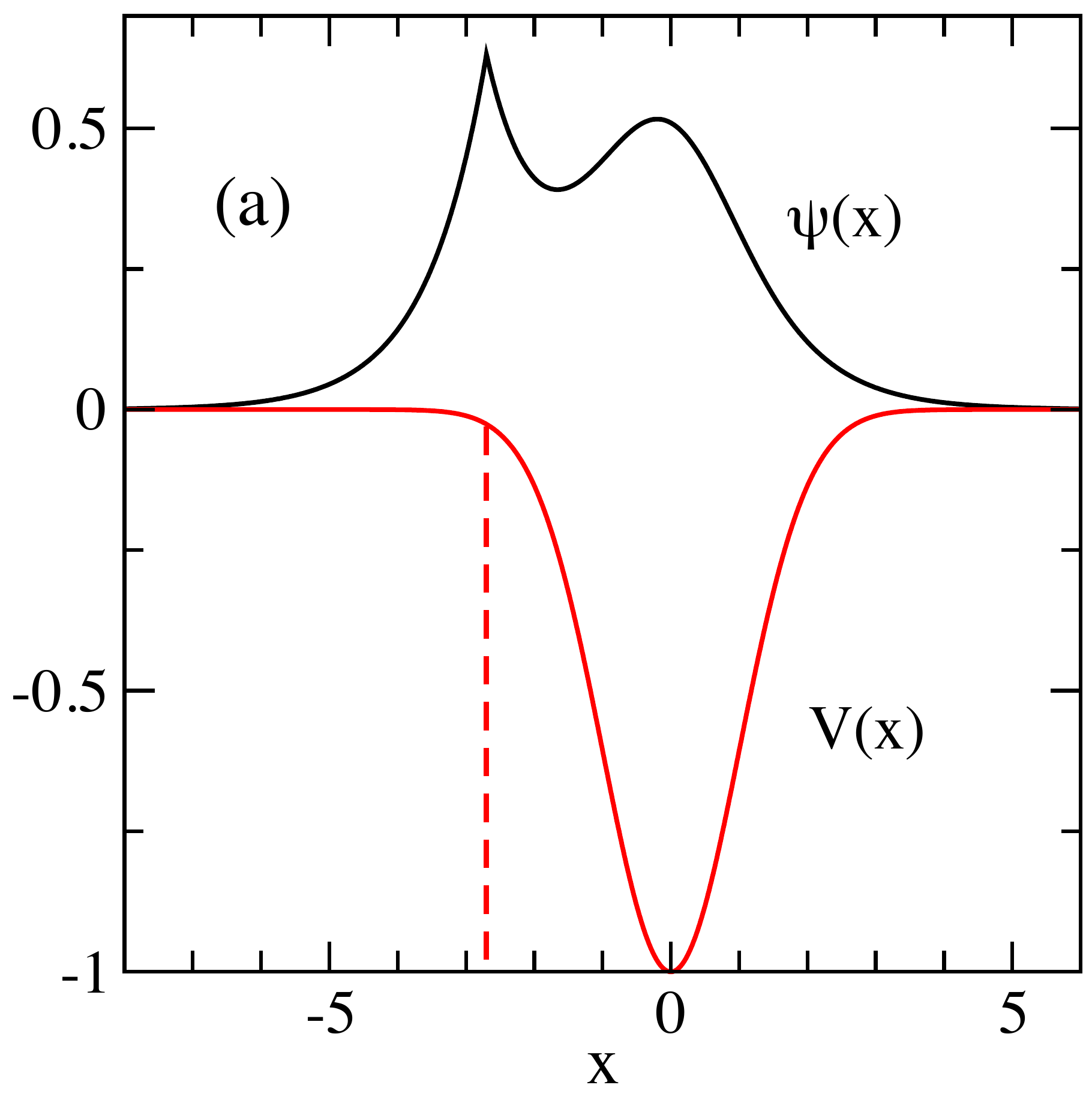}
\includegraphics[width=4.5cm]{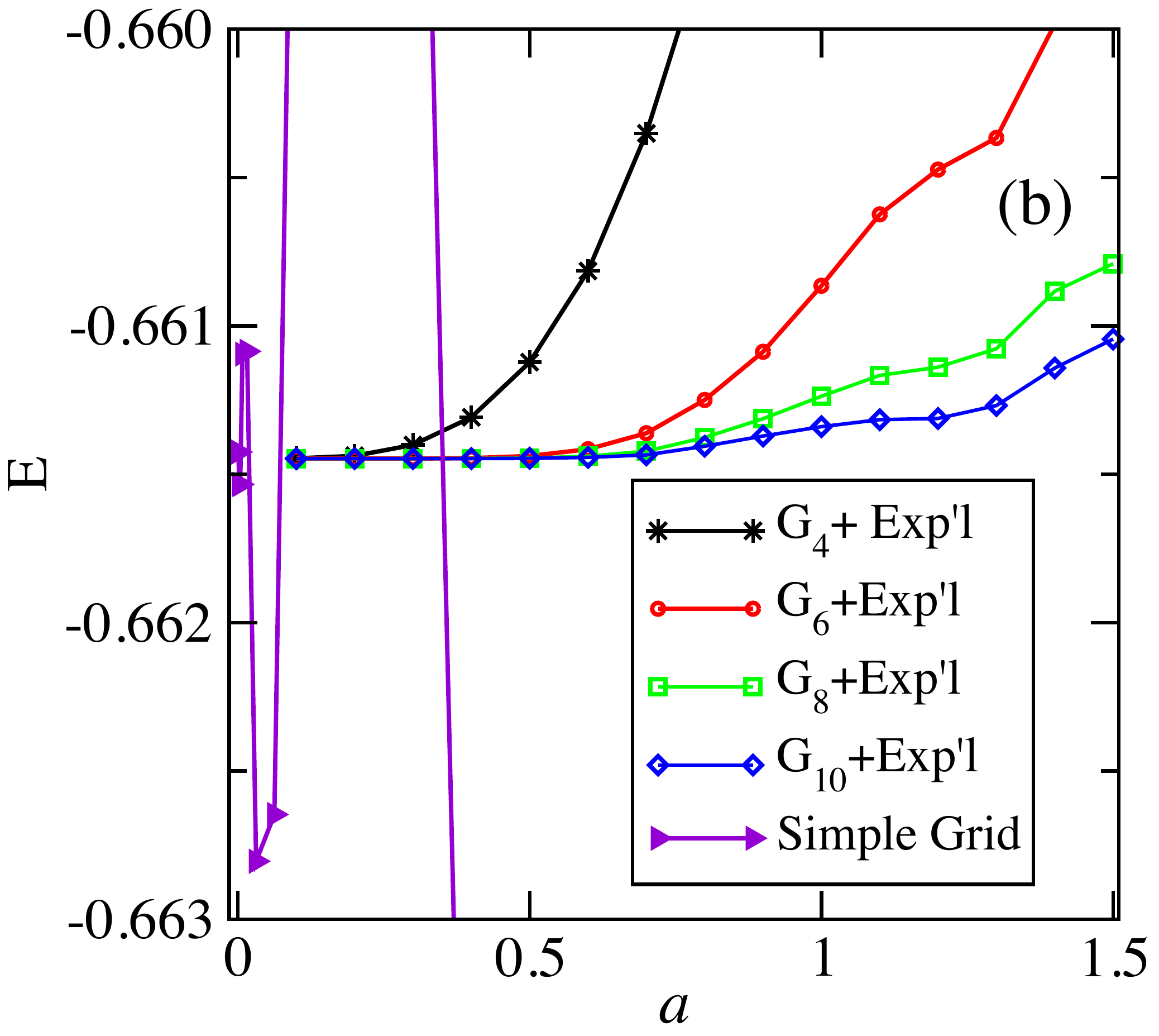} 
\caption{(a) A potential with a smooth part and a delta function, and the resulting ground state. 
(b) Convergence of the energy with grid spacing for two types of bases:  1) an array of gausslets plus an exponential which is the solution of the isolated delta function, for four different gausslets; and 2)
 a simple second order finite difference grid.
}
\label{fig:gaudelta}
\end{figure}

A key advantage of a basis set approach is that one can add functions to it to represent singularities
or other sharp features.  As a test of this, shown in Fig. \ref{fig:gaudelta} we consider a potential with a smooth
part plus an off-center delta function, $V(x) = -\exp(-\frac{1}{2} x^2) -\delta(x-2.7)$. Particularly in three dimensions,
this sort of potential is difficult with a grid. Our basis consists of an array of gausslets plus, to represent the singularity, the solution
to the delta function alone, $\exp(-|x-2.7|)$.  The exponential is orthogonalized to the gausslets to make
a fully orthogonal basis. To carry out the calculations, the exponential is first represented very accurately
as a sum of a few hundred Guassians, the coefficients coming from a discretization of an integral. 
(In a 3D calculation, to describe, say, a 1S function, one would use a sum of a few 
Gaussians from a standard basis set.)
Then all the potential terms are sums of analytic Gaussian integrals.
The hybrid Gaussian/exponential approach converges very rapidly, with errors below about $10^{-3}$ 
for $a < 1.5$ with \Gcalnosp{8} and \Gcalnosp{10}. The energies for the last two points, $a=0.1$ and $a=0.2$, agree to eight digits,
$E = -0.66144716$.  For comparison, we implemented a naive grid, without trying to put a grid
point exactly on the singularity.  The grid results show substantial oscillations as a function of $a$.
The finest grid result, $a = 10^{-4}$, gives $E \approx -0.6614$, in agreement with the gausslet
result but much less accurate.

\section{Diagonal Approximations}

A crucial feature of finite difference grid approximations is
the simple form of the two electron interaction, $V_{ij} \hat n_i \hat n_j$.  A similar property holds for
the one electron nucleus-electron potential terms, which have the simple form $U_i \hat n_i$. 
This latter
property is not so important in itself, since the extra computation time to deal with a $U_{ij}$ form of
the potential is minor, but it does serve another useful purpose: approximations which make $U$ diagonal
should generally do the same for $V$, since the two electron interaction acts on one electron with a single-particle
potential determined by the locations of all the other electrons. We will consider the simpler single particle
potential first, in one dimension.  Diagonal representations such as these are a key property of bases made from
the sinc function\cite{Jones:2016}.  Here we find similar approximations for gausslets, implying that the high
nonlocality of the sinc is not needed for this sort of approximation.  Although we focus on the gausslets, similar
approximations would work for the ternary wavelet scaling functions, and for certain types of traditional wavelet
scaling functions with the $\delta$-function property, particularly coiflets\cite{Daubechies:1992}.

Let $U(x)$ act on a continuum wavefunction $\psi(x)$ to give another wavefunction $\phi(x)$:   $\phi(x) = U(x) \psi(x)$.
We are  interested in the coefficient of $\phi$ for basis function ${\mathcal G}_i$, written using the
expansion of  $\psi$ in terms of the ${\mathcal G}_j$:  
\begin{equation}
    \phi_i = \int dx\  {\mathcal G}_i(x) \phi(x) = \sum_j\psi_j \int dx\  {\mathcal G}_i(x) U(x) {\mathcal G}_j(x) .
\end{equation}
This is the conventional non-diagonal form, with the last integral defining $U_{ij}$.
Now assume that each ${\mathcal G}_i$ integrates like a weighted delta function, with location $x_i$:
\begin{equation}
    \int dx\  {\mathcal G}_i(x) f(x)  = f(x_i) w_i 
\end{equation}
for any smooth function $f(x)$,
with 
\begin{equation}    
    w_i \equiv \int dx\  {\mathcal G}_i(x) .
\end{equation}
Then 
\begin{equation}
\phi_i = \phi(x_i) w_i = U(x_i) \psi(x_i) w_i = U(x_i) \psi_i.
\end{equation}
Thus we have a diagonal ``point'' approximation for $U$:
\begin{equation}
U_{ij}  \to \delta_{ij} U(x_i).
\end{equation}
Note that the locality of the gausslets means that $U_{ij}$ is small for $i$ far from $j$, but the diagonal approximation does not rely on this.  In particular, we disregard the near-neighbor $j=i\pm1$ terms even though they are not small.

There are two other closely related diagonal approximations.  First, if the $\delta$-function property is
only approximate, we may prefer to replace $U(x_i)$ with its overlap with ${\mathcal G}_i$:
\begin{equation}
    U_{ij} \to \delta_{ij}  \int dx\ {\mathcal G}_i(x) U(x) / w_i .\label{USU}
\end{equation}
One might hope that this ``integral'' approximation could be more accurate, 
since it averages the potential over a finite
range, but one needs tests to see if it actually is an improvement.
Second, let us assume that the $\{{\mathcal G}_i\}$ can exactly represent a constant function over the range
of interest.  Then $w_k$ is then the expansion coefficient for ${\mathcal G}_k$ to represent the identity function, and
\begin{equation}
    1 = \sum_k w_k {\mathcal G}_k(x) .
\end{equation}
Inserting this into Eq. (\ref{USU}) gives the ``summed'' approximation
\begin{equation}
U_{ij}  \to \delta_{ij}  \sum_k U_{ik} w_k / w_i . \label{Uik}
\end{equation}
For the special case where the ${\mathcal G}_i$ are a uniform grid of gausslets, $w_k = w_i$, and
as the grid spacing goes to zero,
smooth functions have $\psi_k \approx \psi_i $ in the close vicinity of a point $i$.  In this case,
$\sum_k U_{ik} \psi_k \approx \psi_i \sum_k U_{ik}$, providing another 
rough justification for Eq. (\ref{Uik}).  (This idea was the original inspiration for these diagonal
approximations.)

\begin{figure}[t]
\includegraphics[width=7cm]{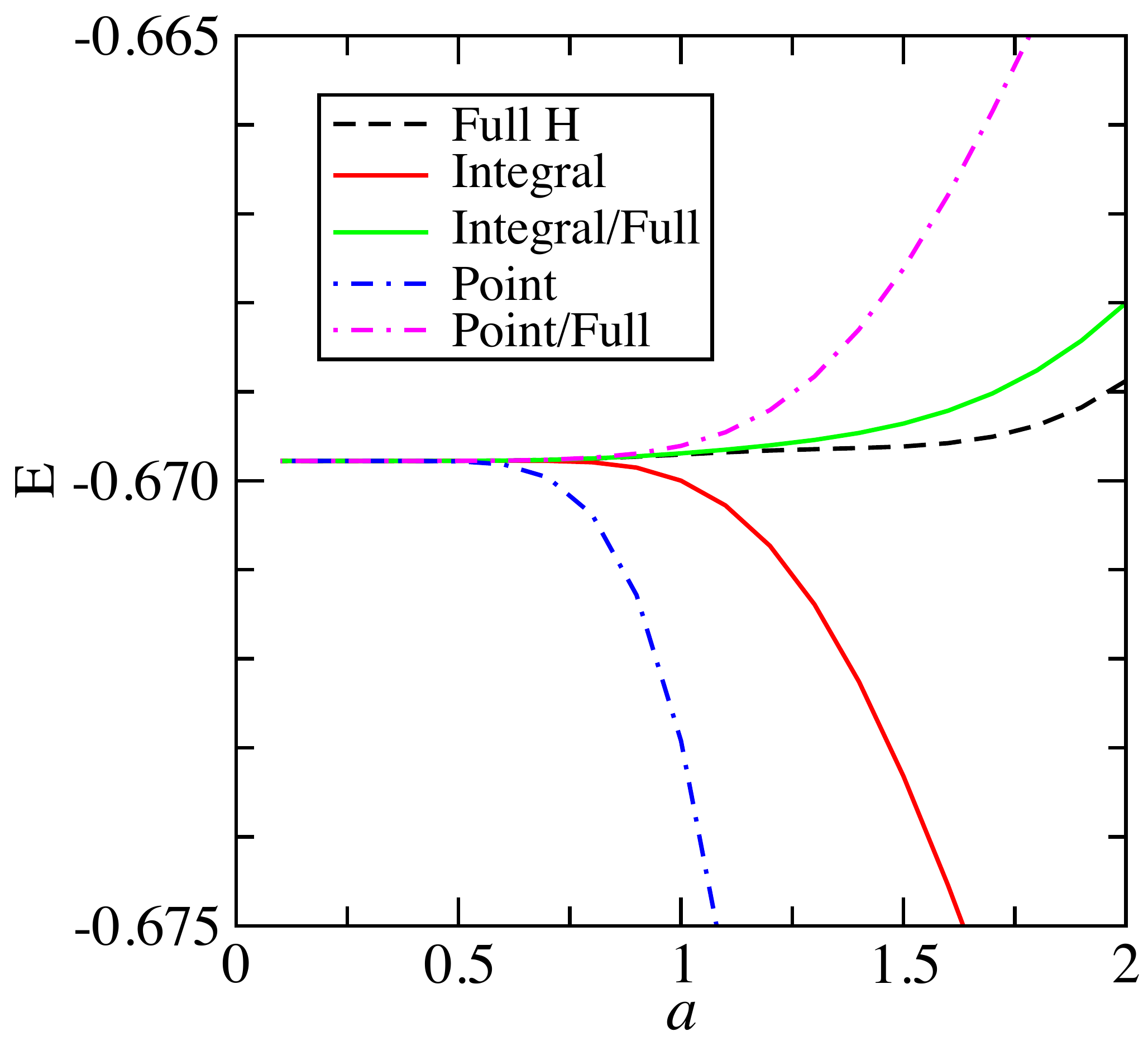}
\caption{Energy solving the soft Coulomb potential (with $b=0$) using diagonal approximations with the \Gcalnosp{10} gausslet basis, as a function of spacing $a$.
The ``Full'' curve uses the full Hamiltonian matrix.  The ``Integral'' diagonal approximation is shown, which for these
bases is identical to the ``Summed'' approximation.  For the two diagonal approximations, we also show the energy evaluated with
the full Hamiltonian, but with the eigenvector from the corresponding approximation.
  }
\label{fig:figB}
\end{figure}

Tests of the diagonal approximations are shown for the soft Coulomb potential in Fig. \ref{fig:figB}.
One sees that all of the diagonal approximations become very precise for small $a$.  In fact, although one cannot see it in this figure, the diagonal approximation curves and the non-diagonal full matrix curve
come together faster as a function of $a$ than any of the curves converge to the exact result. 
This is because the $\delta$-function property paramenter $p$ is twice that of the completeness parameter $l$.
This means that at small $a$, there is no point in using the full matrix--the result might not be exact,
but the diagonal approximation doesn't contribute significantly to any error.  At larger $a$, the diagonal
approximations introduce significant errors.  The integral approximation (or equivalently for this case,
the summed approximation) is significantly better than the point approximation.
 
 These three approximations translate immediately into two electron approximations for $V$.
The pointwise evaluation is
 \begin{equation}
V_{ijkl}  \to \delta_{il} \delta_{jk} V(x_i,x_j) . \label{pointwise}
\end{equation}
The integral approximation replaces $V(x_i,x_j)$ in Eq. (\ref{pointwise}) with
  \begin{equation}
      V(x_i,x_j) \to \int dx\ dx'\ {\mathcal G}_i(x) V(x,x') {\mathcal G}_j(x') / (w_i w_j) .
\end{equation}
The summed approximation replaces $V(x_i,x_j)$ with
  \begin{equation}
V(x_i,x_j) \to \sum_{kl} V_{ijkl} w_l w_k / (w_i w_j)
\end{equation}
Because of the big difference in computational scaling associated with two electron versus
one electron terms, it is sensible to only make the diagonal approximation for the two
electron terms, using the full matrices for the one electron terms.

\begin{figure}[t]
\includegraphics[width=7cm]{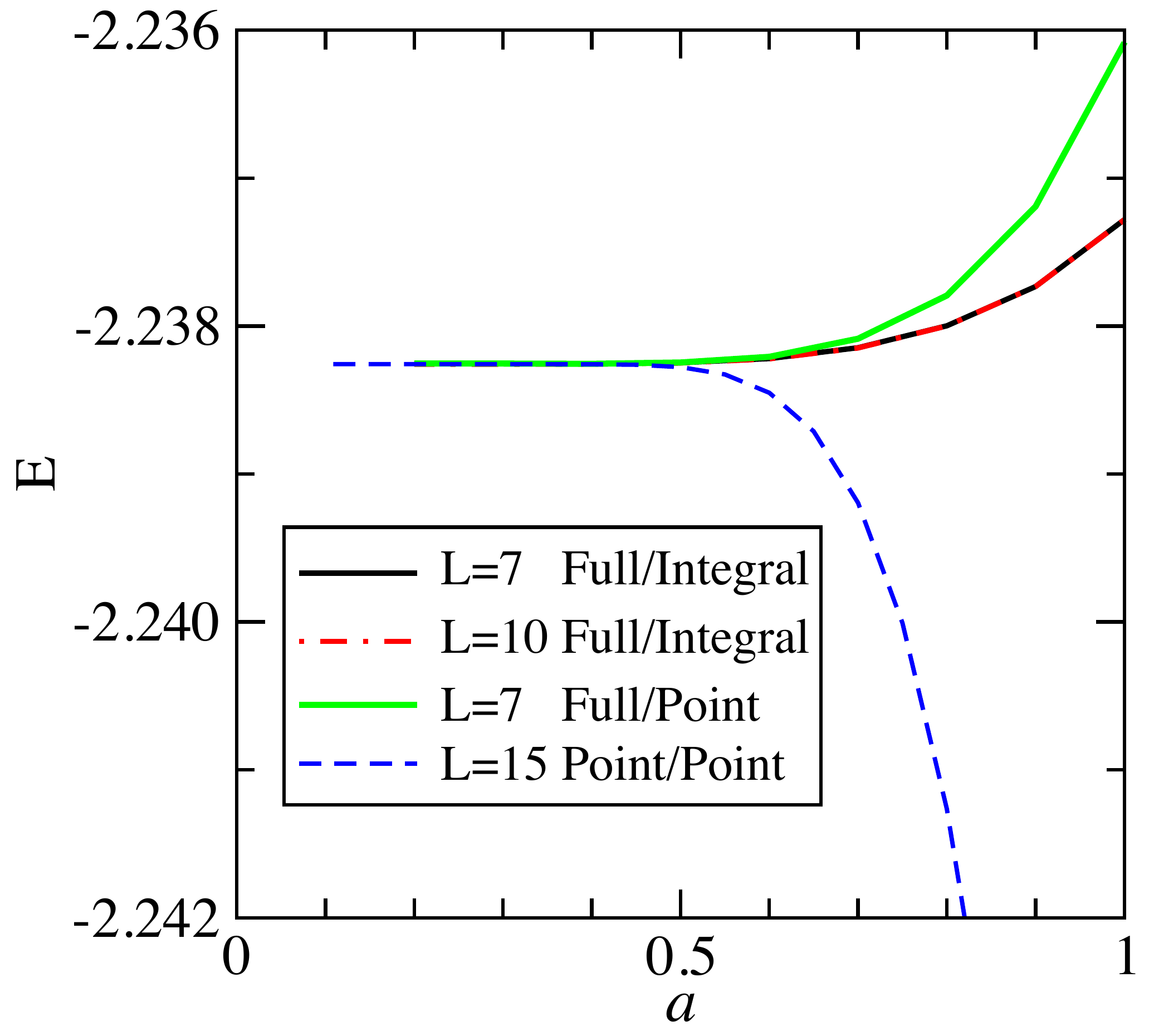}
\caption{Energy solving the soft Coulomb 1D helium atom using gausslet basis \Gcalnosp{10} with diagonal approximations, 
as a function of spacing $a$.
The approximations are labeled in the form (single particle approximation)/(two particle approximation).  The
lattice of gausslets extends approximately from $-L$ to $L$. 
  }
\label{fig:figC}
\end{figure}

As a simple interacting example calculation we perform an exact diagonalization for a 1D helium atom in
a soft Coulomb potential, with a basis of gausslets ranging from $-L$ to $L$.  The results are shown in Fig. \ref{fig:figC}. 
According to Ref. \cite{Wagner:2012}, the exact energy from a grid DMRG
calculation is -2.238.  
The two Full/Integral curves, which completely overlap, 
demonstrate that $L=7$ gives excellent convergence in $L$.
All the diagonal approximations perform excellently at smaller $a$.
For larger $a$ they still behave well; the Full/Integral approximation stays within 1 mH almost
up to $a=1$.
The point diagonal approximations are somewhat less accurate at larger $a$, but the Point/Point approximation is especially
convenient for small $a$, since it is as accurate as the other approximations and requires no integral evaluations at all.
For this case we went to $a=0.1$ and $L=15$, which gave 601 basis functions, to push for very high accuracy; we obtain
$E=-2.238257824$, which we believe is correct to all the digits shown.  The setup and diagonalization took less than
5 minutes on a desktop. With DMRG, it would be straightforward to extend these calculations to long chains.

One can consider making the diagonal approximations after first making a mean-field-like reduction
of the operator.  For the one electron case derived below, this gives an energy correction but does
not change the Hamiltonian.  For the two electron terms, this produces altered one electron
terms which change the diagonalization and might improve the results.  For the one electron
case, assume we have some estimate for $\langle \hat c^\dagger_i \hat c_j \rangle$, and let
\begin{eqnarray}
\hat U &=& \sum_{ij} U_{ij} \hat c^\dagger_i \hat c_j\\
&=& \sum_{ij} U_{ij} (\hat c^\dagger_i \hat c_j - \langle \hat c^\dagger_i \hat c_j \rangle) +
\sum_{ij} U_{ij}  \langle \hat c^\dagger_i \hat c_j \rangle .
\end{eqnarray}
If we now apply a diagonal approximation $U_{ij} \to \tilde U_{ii}$ on the first term, we have
 \begin{equation}
     \hat U \approx \sum_{i} \tilde U_{ii} \hat c^\dagger_i \hat c_i + \left( 
\sum_{ij} U_{ij} \langle \hat c^\dagger_i \hat c_j \rangle -
     \sum_{i} \tilde U_{ii} \langle \hat n_i\rangle\right).
     \label{meanfield}
\end{equation}
Suppose we use as our approximate $\langle \hat c^\dagger_i \hat c_j \rangle$ (and $\langle \hat n_i \rangle$)
the result from
the ground state using $\tilde U$. Then adding 
the correction term in the parentheses in Eq. (\ref{meanfield})
is the same as evaluating the energy in the full Hamiltonian of the eigenstate from the approximate Hamiltonian.
This corrected result is shown in Fig. \ref{fig:figB}. We see that the corrected result is significantly
better than the uncorrected result at large $a$.  We leave exploring these mean-field
approximations for the two electron term for future work.

A basis permitting a good diagonal approximation is special, and an orthogonal transformation 
of the basis will, in general, spoil the approximation.  A truncation of the basis may not
harm the approximation; in particular, a reduction of a particular set of orbitals into {\it one}    
orbital which is a linear combination from the set still allows a diagonal approximation.
This is because a basis rotation takes two particle fermion number-conserving operators 
(i.e. $\hat c^\dagger_i \hat c_j$ or $\hat n_i$) into similar
two particle operators, and a single orbital only has one such operator, $\hat n_i$.
This applies also to the two-electron terms.
If one knows the single particle reduced density matrix (RDM, $\langle \hat c^\dagger_i \hat c_j \rangle$), 
one can determine if reducing a particular set of functions to one function is a good approximation:
it corresponds to the diagonal block over the sites of the RDM having just one significant eigenvalue.
If so, that eigenvector would be the correct linear combination.
One may expect that the outer tail regions of a molecule would satisfy this criterion, 
since far from the nuclei, the wavefunction decays in a simple way.
This suggests that one should combine sites in the tail regions into tail functions, which
could lead to a large reduction in the number of basis functions.
If one transforms a set of basis function into a few functions, this breaks the diagonal
approximation, but it only generates a limited number of off-diagonal terms, which may be
acceptable.  Specifically, if basis function $i$ is in the reduced set and $p$, $q$, and $r$ are outside the set,
terms such as $\hat c^\dagger_i \hat c^\dagger_p c_q c_r$) with only one (or three) operators 
in the set are not generated.

A very important question is how to make a diagonal or near-diagonal approximation with a basis composed
of an array of gausslets with some additional non-gausslet functions, particularly 
to represent the sharp behaviour near nuclei.  We leave this question for future studies.

\section{Discussion}
Application of these ideas to sliced basis DMRG would require little development beyond what is presented here.
The initial applications of SBDMRG used a finite difference grid which was adapted with a low-frequency 
filtering procedure to increase convergence with the grid spacing.  
Usually a grid spacing of 0.1 Bohr was used for hydrogen chains.  The grid plus
filtering procedure could be replaced by a gausslet basis, with a diagonal approximation introduced for the
two-electron interaction.  Probably this would allow substantially higher lattice spacing, and it would smooth
the way for including larger $Z$ atoms.

More generally, for realistic 3D systems,  one can form a 3D gausslet basis by making products of 1D gausslets, namely
\begin{equation}
    {\mathcal G}(x,y,z) = {\mathcal G}(x)  {\mathcal G}(y)  {\mathcal G}(z)  
\end{equation}
If one is not worried about making diagonal approximations, then one can combine gausslets with 
3D Gaussians from a standard Gaussian basis, orthogonalizing
them to the gausslets. 
The fact that gausslets are made of Gaussians would facilitate this.  
Note that for a uniform grid of gausslets, all the two electron integrals between gausslets only
need be done only once,
and would be applicable to all systems, with a simple rescaling for different lattice spacings.

Diagonal approximation can be extremely effective in speeding up calculations, and so one
would want to try to find approaches that allow them.
The most straightforward approach would involve using a pseudopotential both of the usual 
one-electron type, and also a two-electron pseudopotential\cite{Prendergast:2001}.  This one electron pseudopotential 
means that a gausslet 3D grid, without supplementary Gaussians, would be adequate, 
and the two electron pseudopotental would allow any of our diagonal approximationn for the two electron
interaction. The two-electron diagonal approximation would greatly speed up DMRG, and also allow other tensor network
methods, such as projected entangled pair states (PEPS) to be tried. 
The diagonal interaction could be compressed for use in DMRG, just as is done in SBDMRG.
The diagonal approximation may also speed up other correlation methods.
To our knowledge, the combination of both types of pseudopotentials has not been used before, but it
seems reasonably straightforward.  The key question would be:  what grid spacing is needed for reasonable
accuracy?  

For all electron calculations combining gausslets and 3D Gaussians, 
two-electron diagonal approximations would need to be tested and developed, since the extra functions do not
fit within the $\delta$-function framework we used to derive diagonal approximations. Nevertheless,
one might find approximations with acceptable accuracy.
One could also
consider whether diagonal approximations could be useful even in a standard Gaussian basis which has
been localized by a standard method. Recently, Baker, Burke, and White\cite{Baker:2017} have proposed ``wavelet localization'' (WL), where an auxiliary wavelet basis is used to localize an existing delocalized basis, at the cost of a modest increase in the number of functions in the basis. One could equally use gausslets to perform wavelet localization.  One scheme would be to WL to localize standard Gaussian atom-centered basis functions on each atom. The result would be
functions which look standard Gaussians close to their nucleus, but midway to the next atom they would rapidly die off with oscillations to make them orthogonal to all functions on neighboring atoms. This might give rise to a modified diagonal approximation in which $V_{ijkl}$ is nonzero only if $i$ and $l$ are on the same atom, and $j$ and $k$ are on the same atom. 


In conclusion, we have introduced gausslets, a new type of basis function, which combine the
efficiencies of working with Gaussians, but with systematic completeness and orthogonality, while
maintaining locality, symmetry, and smoothness.  We have
introduced diagonal approximations, which are tied to gausslets, 
which dramatically improve the scaling of electronic structure
calculations, making them act more like a grid than a basis. 
Although the various tests we have performed here are in 1D, these basis functions were
developed with 3D applications in mind, and we anticipate rapid development of 3D uses.

We thank Miles Stoudenmire, Glen Evenbly, Kieron Burke, Takeshi Yanai, Tom Baker,
and Garnet Chan for  helpful conversations.
We acknowledge support from the Simons Foundation through the Many-Electron Collaboration,
and from the U.S. Department of
Energy, Office of Science, Basic Energy Sciences under
award \#DE-SC008696.

\bibliography{paper}


\section{Appendix:  Coefficients of wavelets and gausslets}

\setcounter{totalnumber}{8}
\setcounter{topnumber}{8}
\setcounter{bottomnumber}{8}

\renewcommand{\topfraction}{2.0}
\renewcommand{\bottomfraction}{2.0}
\renewcommand{\textfraction}{0.0}
\renewcommand{\floatpagefraction}{0.0}

Here we give coefficients for the wavelets and gausslets described here.
First, we present the coefficients of the gausslets, which are defined by
\begin{equation}
    {\mathcal G}(x) = \sum_j b_j \exp[-\frac{1}{2}(3x-j)^2]   \label{Gbj}
\end{equation}
 The $b_j$ for gausslets with even order 4 through 10 are given in Tables II-V.
 
 In Tables VI-XI, the coefficients defining ternary wavelet transforms are given.
 These can be used to define sequence of gausslets based on the primary gausslets
 given in Tables II-V.  For example, applying a WT to a primary gausslet gives a
 gausslet with Gaussian spacing of $1/9$. Combining Eqs. (\ref{Gbj}) and (\ref{ckf}), we have
 \begin{equation}
    {\mathcal G'}(x) = \sum_l \exp[-\frac{1}{2}(9x-l)^2] \sum_k c_k b_{l-3k},  \label{Gpbj}
\end{equation}
where the $b_j$ come from the primary gausslet and the $c_k$ are the coefficients associated with
the additional WT.  To define wavelet-like functions coming from this transformation, we use the same
formula, but with the mid or high wavelet coefficients replacing the $c_k$.  One can repeat these WTs to
construct a multi-level wavelet-like basis, where the functions at each scale, in addition to different scalings, are slightly different.  All the basis functions in the multi-level basis would be written in terms a single grid of Gaussians, which would have a spacing a factor of three smaller than  the finest level of wavelet-like functions.

\begin{table*}[p]
    \begin{tabular}{|c|r||c|r||c|r||c|r|}
	\hline
	$j$ & $b_j$ \qquad \qquad \qquad & $j$ & $b_j$\qquad \qquad \qquad &
	$j$ & $b_j$ \qquad \qquad \qquad & $j$ & $b_j$\qquad \qquad \qquad \\ 
	\hline
 $0$& $  0.6067686239029718$  & $12$& $ -0.0007634581491839$  & $24$& $ -0.0000000111857638$  & $36$& $ -0.0000000000000001$  \\
 $1$& $  0.4595762731397992$  & $13$& $  0.0001654718546793$  & $25$& $  0.0000000057570184$  & $37$& $  0.0000000000000000$  \\
 $2$& $ -0.0164427190204405$  & $14$& $ -0.0001062974603036$  & $26$& $  0.0000000017544093$  & $38$& $  0.0000000000000003$  \\
 $3$& $ -0.1403618655345050$  & $15$& $  0.0001489730491976$  & $27$& $ -0.0000000078396493$  & $39$& $  0.0000000000000004$  \\
 $4$& $ -0.0402939401786279$  & $16$& $ -0.0000925527267679$  & $28$& $  0.0000000029209953$  & $40$& $  0.0000000000000003$  \\
 $5$& $  0.0091345923715139$  & $17$& $  0.0000589148129012$  & $29$& $ -0.0000000011159594$  & $41$& $ -0.0000000000000004$  \\
 $6$& $  0.0412074716875908$  & $18$& $ -0.0000343962098581$  & $30$& $  0.0000000006646346$  & $42$& $ -0.0000000000000005$  \\
 $7$& $ -0.0263104231001814$  & $19$& $  0.0000139465804941$  & $31$& $ -0.0000000001080798$  & $43$& $ -0.0000000000000004$  \\
 $8$& $  0.0120390822107673$  & $20$& $ -0.0000055900603412$  & $32$& $ -0.0000000000000000$  & $44$& $ -0.0000000000000001$  \\
 $9$& $ -0.0041120776084794$  & $21$& $  0.0000021993752470$  & $33$& $ -0.0000000000000000$  & $45$& $ -0.0000000000000001$  \\
 $10$& $  0.0003155814182348$  & $22$& $ -0.0000007773157565$  & $34$& $ -0.0000000000000000$  & $46$& $ -0.0000000000000001$  \\
 $11$& $  0.0008905265359326$  & $23$& $  0.0000001821311458$  & $35$& $  0.0000000000000000$  & $47$& $  0.0000000000000000$  \\
 $12$& $ -0.0007634581491839$  & $24$& $ -0.0000000111857638$  & $36$& $ -0.0000000000000001$  & $48$& $  0.0000000000000001$  \\

	\hline
    \end{tabular}
    \caption{Coefficients $b_j$ of the Gaussians defining the gausslet \Gcalnosp{4} for $j\ge 0$, where $b_{-j}=b_j$.
    }
    \label{table:bj4}
\end{table*}

\begin{table*}[p]
    \begin{tabular}{|c|r||c|r||c|r||c|r|}
	\hline
	$j$ & $b_j$ \qquad \qquad \qquad & $j$ & $b_j$\qquad \qquad \qquad &
	$j$ & $b_j$ \qquad \qquad \qquad & $j$ & $b_j$\qquad \qquad \qquad \\ 
	\hline
 $0$& $  0.6510799122138565$  & $10$& $ -0.0129995132051085$  & $20$& $  0.0000527579539840$  & $30$& $  0.0000000012233441$  \\
 $1$& $  0.3748901951337270$  & $11$& $  0.0083444621145336$  & $21$& $ -0.0000284802656230$  & $31$& $  0.0000000063863555$  \\
 $2$& $  0.0939399437214329$  & $12$& $ -0.0035602045266604$  & $22$& $  0.0000150015015272$  & $32$& $ -0.0000000030684215$  \\
 $3$& $ -0.1569006465627569$  & $13$& $  0.0012544959501549$  & $23$& $ -0.0000068808321161$  & $33$& $  0.0000000004500457$  \\
 $4$& $ -0.0948155527751206$  & $14$& $  0.0003594627655807$  & $24$& $  0.0000028004555091$  & $34$& $ -0.0000000002218040$  \\
 $5$& $  0.0232646256086860$  & $15$& $ -0.0007848423809006$  & $25$& $ -0.0000013067346743$  & $35$& $  0.0000000000913882$  \\
 $6$& $  0.0216613768304792$  & $16$& $  0.0005747148102592$  & $26$& $  0.0000007168554123$  & $36$& $  0.0000000000104569$  \\
 $7$& $  0.0361805021062946$  & $17$& $ -0.0003070053297971$  & $27$& $ -0.0000003912054599$  & $37$& $ -0.0000000000043992$  \\
 $8$& $ -0.0317148981502408$  & $18$& $  0.0001499886588480$  & $28$& $  0.0000001613546686$  & $38$& $ -0.0000000000000000$  \\
 $9$& $  0.0133915814065059$  & $19$& $ -0.0000895654658735$  & $29$& $ -0.0000000401650166$  & $39$& $ -0.0000000000000000$  \\
 $10$& $ -0.0129995132051085$  & $20$& $  0.0000527579539840$  & $30$& $  0.0000000012233441$  & $40$& $ -0.0000000000000000$  \\
	\hline
    \end{tabular}
    \caption{Coefficients of the Gaussians defining the gausslet \Gcalnosp{6}.
    }
    \label{table:bj6}
\end{table*}

\begin{table*}[!]
    \begin{tabular}{|c|r||c|r||c|r||c|r|}
	\hline
	$j$ & $b_j$ \qquad \qquad \qquad & $j$ & $b_j$\qquad \qquad \qquad &
	$j$ & $b_j$ \qquad \qquad \qquad & $j$ & $b_j$\qquad \qquad \qquad \\ 
	\hline
 $0$& $  0.6188489361270065$  & $12$& $ -0.0046547275254753$  & $24$& $  0.0000289020341125$  & $36$& $  0.0000000067016738$  \\
 $1$& $  0.3824167454273702$  & $13$& $  0.0040100201083541$  & $25$& $ -0.0000168829682435$  & $37$& $ -0.0000000032075046$  \\
 $2$& $  0.1099474897465580$  & $14$& $ -0.0015184700034549$  & $26$& $  0.0000097119084642$  & $38$& $  0.0000000009153639$  \\
 $3$& $ -0.1478654707279702$  & $15$& $ -0.0002473520884477$  & $27$& $ -0.0000054294948468$  & $39$& $ -0.0000000001508133$  \\
 $4$& $ -0.1092533175894797$  & $16$& $  0.0004021212441880$  & $28$& $  0.0000027905622899$  & $40$& $  0.0000000000575772$  \\
 $5$& $  0.0008350876805188$  & $17$& $ -0.0006709855121816$  & $29$& $ -0.0000011656692668$  & $41$& $ -0.0000000000147799$  \\
 $6$& $  0.0383468513752624$  & $18$& $  0.0006479929834526$  & $30$& $  0.0000004251304827$  & $42$& $ -0.0000000000041069$  \\
 $7$& $  0.0443793867348271$  & $19$& $ -0.0004072275423943$  & $31$& $ -0.0000001918178974$  & $43$& $  0.0000000000015572$  \\
 $8$& $ -0.0264220705098279$  & $20$& $  0.0002507059396952$  & $32$& $  0.0000001011853914$  & $44$& $  0.0000000000000007$  \\
 $9$& $  0.0039445390490703$  & $21$& $ -0.0001490130367618$  & $33$& $ -0.0000000521084195$  & $45$& $  0.0000000000000002$  \\
 $10$& $ -0.0180915921775044$  & $22$& $  0.0000852936240908$  & $34$& $  0.0000000266388167$  & $46$& $  0.0000000000000001$  \\
 $11$& $  0.0130345989975900$  & $23$& $ -0.0000498804314109$  & $35$& $ -0.0000000129272754$  & $47$& $  0.0000000000000002$  \\
 $12$& $ -0.0046547275254753$  & $24$& $  0.0000289020341125$  & $36$& $  0.0000000067016738$  & $48$& $  0.0000000000000000$  \\
	\hline
    \end{tabular}
    \caption{Coefficients of the Gaussians defining the gausslet \Gcalnosp{8}.
    }
    \label{table:bj8}
\end{table*}

\begin{table*}[!]
    \begin{tabular}{|c|r||c|r||c|r||c|r|}
	\hline
	$j$ & $b_j$ \qquad \qquad \qquad & $j$ & $b_j$\qquad \qquad \qquad &
	$j$ & $b_j$ \qquad \qquad \qquad & $j$ & $b_j$\qquad \qquad \qquad \\ 
	\hline
 $0$& $  0.6006282292783031$  & $17$& $ -0.0009695161114260$  & $34$& $  0.0000004139991910$  & $51$& $ -0.0000000000000030$  \\
 $1$& $  0.3870904132059249$  & $18$& $  0.0012381620748654$  & $35$& $ -0.0000001869527576$  & $52$& $ -0.0000000000000088$  \\
 $2$& $  0.1167436095101837$  & $19$& $ -0.0008657512270795$  & $36$& $  0.0000000787310449$  & $53$& $  0.0000000000000063$  \\
 $3$& $ -0.1401141978512072$  & $20$& $  0.0007050590750442$  & $37$& $ -0.0000000360812189$  & $54$& $ -0.0000000000000019$  \\
 $4$& $ -0.1178552983794614$  & $21$& $ -0.0005322979066705$  & $38$& $  0.0000000168525628$  & $55$& $  0.0000000000000030$  \\
 $5$& $ -0.0112632618094700$  & $22$& $  0.0003332874495659$  & $39$& $ -0.0000000087040883$  & $56$& $ -0.0000000000000018$  \\
 $6$& $  0.0450560144981757$  & $23$& $ -0.0002178032104139$  & $40$& $  0.0000000042255242$  & $57$& $  0.0000000000000002$  \\
 $7$& $  0.0502131666992306$  & $24$& $  0.0001389608411184$  & $41$& $ -0.0000000014639246$  & $58$& $ -0.0000000000000001$  \\
 $8$& $ -0.0207372799495982$  & $25$& $ -0.0000849543923289$  & $42$& $  0.0000000006473451$  & $59$& $ -0.0000000000000004$  \\
 $9$& $ -0.0031814624464224$  & $26$& $  0.0000533515010750$  & $43$& $ -0.0000000003739856$  & $60$& $  0.0000000000000005$  \\
 $10$& $ -0.0214900136942583$  & $27$& $ -0.0000327971054166$  & $44$& $  0.0000000001419863$  & $61$& $ -0.0000000000000004$  \\
 $11$& $  0.0139369308627208$  & $28$& $  0.0000193278214075$  & $45$& $ -0.0000000000705564$  & $62$& $  0.0000000000000003$  \\
 $12$& $ -0.0029594340072233$  & $29$& $ -0.0000108674604171$  & $46$& $  0.0000000000418054$  & $63$& $ -0.0000000000000003$  \\
 $13$& $  0.0057046712233152$  & $30$& $  0.0000060213353043$  & $47$& $ -0.0000000000097639$  & $64$& $  0.0000000000000001$  \\
 $14$& $ -0.0026819334185882$  & $31$& $ -0.0000033140396282$  & $48$& $ -0.0000000000008755$  & $65$& $ -0.0000000000000001$  \\
 $15$& $ -0.0004611902203357$  & $32$& $  0.0000016801358258$  & $49$& $  0.0000000000004776$  & $66$& $  0.0000000000000000$  \\
 $16$& $  0.0003205662299202$  & $33$& $ -0.0000008242534887$  & $50$& $ -0.0000000000000061$  & $67$& $ -0.0000000000000001$  \\
 $17$& $ -0.0009695161114260$  & $34$& $  0.0000004139991910$  & $51$& $ -0.0000000000000030$  & $68$& $  0.0000000000000000$  \\
	\hline
    \end{tabular}
    \caption{Coefficients of the Gaussians defining the gausslet \Gcalnosp{10}.
    }
    \label{table:bj10}
\end{table*}

\begin{table*}[t!]
    \begin{tabular}{|c||r|r|r|}
	\hline
	$i$ & $c_i$ \qquad \qquad \qquad & mid \quad \quad \quad & high\qquad \qquad \qquad \\ 
	\hline
$ 0$ &$   0.7484170156161815$ &$ -0.4141800601853688$ &$ -0.2024957614263287$ \\
$ 1$ &$   0.4276668660663895$ &$  0.5561945933948288$ &$  0.6771570498285660$ \\
$ 2$ &$   0.1710667464265558$ &$  0.5561945933948288$ &$ -0.6771570498285660$ \\
$ 3$ &$  -0.0855333732132779$ &$ -0.4141800601853688$ &$  0.2024957614263287$ \\
$ 4$ &$  -0.0213833433033195$ &$ -0.1375657059183308$ &$  0.0166032494845936$ \\
$ 5$ &  $-$\qquad\qquad\qquad  &$ -0.0118635394430115$ &$ -0.0118635394430115$ \\
$ 6$ &  $-$\qquad\qquad\qquad  &$  0.0059317697215057$ &$  0.0059317697215057$ \\
$ 7$ &  $-$\qquad\qquad\qquad  &$  0.0014829424303764$ &$  0.0014829424303764$ \\
	\hline
    \end{tabular}
    \caption{Coefficients of wavelet transform $W_{220}$, for the scaling function $c_i$, and the
	middle and high momentum wavelets, for $i \ge 0$. The scaling function is even, whereas the other
	two are even or odd about $i=1.5$.
    }
    \label{table:c220}
\end{table*}

\begin{table*}[t]
    \begin{tabular}{|c||r|r|r|}
	\hline
	$i$ & $c_i$ \qquad \qquad \qquad & mid \quad \quad \quad & high\qquad \qquad \qquad \\ 
	\hline
$ 0$ &$   0.6969234250586759$ &$ -0.4107941294958309$ &$ -0.2912209736267808$ \\
$ 1$ &$   0.4927992798267444$ &$  0.5569318768365513$ &$  0.6414828661994326$ \\
$ 2$ &$   0.1020620726159658$ &$  0.5569318768365513$ &$ -0.6414828661994326$ \\
$ 3$ &$  -0.0597865779345251$ &$ -0.4107941294958309$ &$  0.2912209736267808$ \\
$ 4$ &$  -0.0175110832530844$ &$ -0.1450832262860604$ &$  0.0605322369231791$ \\
$ 5$ &  $-$\qquad\qquad\qquad  &$ -0.0043460190752818$ &$ -0.0043460190752818$ \\
$ 6$ &  $-$\qquad\qquad\qquad  &$  0.0025458390319679$ &$  0.0025458390319679$ \\
$ 7$ &  $-$\qquad\qquad\qquad  &$  0.0007456589886540$ &$  0.0007456589886540$ \\
	\hline
    \end{tabular}
    \caption{Coefficients as above for $W_{212}$.
    }
    \label{table:c212}
\end{table*}
\begin{table*}[]
    \begin{tabular}{|c||r|r|r|}
	\hline
	$i$ & $c_i$ \qquad \qquad \qquad & mid \quad \quad \quad & high\qquad \qquad \qquad \\ 
	\hline
$ 0$ &$   0.6654518154646826$ &$ -0.4617717901799725$ &$ -0.2359874179084521$ \\
$ 1$ &$   0.4969889091605378$ &$  0.5186541524715275$ &$  0.6585998426666274$ \\
$ 2$ &$   0.1522417442806671$ &$  0.5186541524715275$ &$ -0.6585998426666274$ \\
$ 3$ &$  -0.0593919977778424$ &$ -0.4617717901799725$ &$  0.2359874179084521$ \\
$ 4$ &$  -0.0672354496530670$ &$ -0.1078637175407073$ &$  0.0841607095236212$ \\
$ 5$ &$  -0.0093917416781021$ &$ -0.0397145690376490$ &$  0.0209204101741108$ \\
$ 6$ &$   0.0157358047969966$ &$  0.0557630905226228$ &$ -0.0526670507235282$ \\
$ 7$ &$   0.0052021123443088$ &$  0.0378621556799923$ &$ -0.0160908994076716$ \\
$ 8$ &$   0.0001324480078012$ &$  0.0011113995954888$ &$  0.0003302461490658$ \\
$ 9$ &$  -0.0003945801566827$ &$ -0.0022730422083537$ &$  0.0000541173646729$ \\
$ 10$ &$  -0.0005877532725198$ &$ -0.0018397302114096$ &$  0.0016267280773742$ \\
$ 11$ &  $-$\qquad\qquad\qquad  &$ -0.0000112285715763$ &$ -0.0000112285715763$ \\
$ 12$ &  $-$\qquad\qquad\qquad  &$  0.0000334514018403$ &$  0.0000334514018403$ \\
$ 13$ &  $-$\qquad\qquad\qquad  &$  0.0000498280781966$ &$  0.0000498280781966$ \\
	\hline
    \end{tabular}
    \caption{Coefficients as above for $W_{432}$.
    }
    \label{table:c432}
\end{table*}
\begin{table*}[]
    \begin{tabular}{|c||r|r|r|}
	\hline
	$i$ & $c_i$ \qquad \qquad \qquad & mid \quad \quad \quad & high\qquad \qquad \qquad \\ 
	\hline
$ 0$ &$   0.6501872253871461$ &$ -0.3326691320361256$ &$ -0.3985523539040441$ \\
$ 1$ &$   0.4940964120768750$ &$  0.4617130598294811$ &$  0.5294225272694084$ \\
$ 2$ &$   0.1780984528432653$ &$  0.4617130598294811$ &$ -0.5294225272694084$ \\
$ 3$ &$  -0.0550234989262552$ &$ -0.3326691320361256$ &$  0.3985523539040441$ \\
$ 4$ &$  -0.0909186892745402$ &$ -0.3513429148166311$ &$ -0.1101656407319533$ \\
$ 5$ &$  -0.0223292656979585$ &$  0.2123841660468002$ &$ -0.1722908299587890$ \\
$ 6$ &$   0.0228722887575782$ &$ -0.0435059280158056$ &$  0.1120830232328460$ \\
$ 7$ &$   0.0207692102073512$ &$  0.0471740159564527$ &$  0.0641734780408469$ \\
$ 8$ &$   0.0001875552900496$ &$  0.0472649765483601$ &$ -0.0474205851314999$ \\
$ 9$ &$  -0.0046673963657314$ &$ -0.0321789874755936$ &$  0.0060694381298741$ \\
$ 10$ &$  -0.0029827472651827$ &$ -0.0141320492125154$ &$ -0.0041510125101285$ \\
$ 11$ &$   0.0000122804325728$ &$  0.0038045472726876$ &$ -0.0069020341086599$ \\
$ 12$ &$   0.0004958110781867$ &$  0.0005097996313745$ &$  0.0038288991079757$ \\
$ 13$ &$   0.0002506428079038$ &$  0.0008168646303215$ &$  0.0011331890760763$ \\
$ 14$ &$   0.0001904530104763$ &$  0.0007056901707920$ &$ -0.0000758169804573$ \\
$ 15$ &$  -0.0000956826425385$ &$ -0.0004157014804255$ &$ -0.0000230761855707$ \\
$ 16$ &$  -0.0000240352411869$ &$ -0.0001197879867532$ &$ -0.0000211614918367$ \\
$ 17$ &  $-$\qquad\qquad\qquad  &$ -0.0000232066649304$ &$ -0.0000232066649304$ \\
$ 18$ &  $-$\qquad\qquad\qquad  &$  0.0000116589127129$ &$  0.0000116589127129$ \\
$ 19$ &  $-$\qquad\qquad\qquad  &$  0.0000029286897978$ &$  0.0000029286897978$ \\
	\hline
    \end{tabular}
    \caption{Coefficients as above for $W_{652}$.
    }
    \label{table:c652}
\end{table*}
\begin{table*}[]
    \begin{tabular}{|c||r|r|r|}
	\hline
	$i$ & $c_i$ \qquad \qquad \qquad & mid \quad \quad \quad & high\qquad \qquad \qquad \\ 
	\hline
$ 0$ &$   0.6405114155921802$ &$ -0.3375139980479778$ &$ -0.4039942834789112$ \\
$ 1$ &$   0.4939883211006996$ &$  0.4503853137136165$ &$  0.5230712815513074$ \\
$ 2$ &$   0.1883467419220768$ &$  0.4503853137136165$ &$ -0.5230712815513074$ \\
$ 3$ &$  -0.0509642509914217$ &$ -0.3375139980479778$ &$  0.4039942834789112$ \\
$ 4$ &$  -0.1002087947361240$ &$ -0.3500934773100930$ &$ -0.1039860497525889$ \\
$ 5$ &$  -0.0319876829488389$ &$  0.2262032106618946$ &$ -0.1823451665854768$ \\
$ 6$ &$   0.0264267798480466$ &$ -0.0457081535644925$ &$  0.1059004557185995$ \\
$ 7$ &$   0.0296416130059931$ &$  0.0684899632270586$ &$  0.0775684132075941$ \\
$ 8$ &$   0.0032279612625169$ &$  0.0347100961359410$ &$ -0.0410249000273447$ \\
$ 9$ &$  -0.0085182815554429$ &$ -0.0313871506443303$ &$ -0.0049510556835877$ \\
$ 10$ &$  -0.0068462489140611$ &$ -0.0213667045463818$ &$ -0.0127701397211819$ \\
$ 11$ &$   0.0001096044668196$ &$ -0.0034628183480061$ &$  0.0003939228713510$ \\
$ 12$ &$   0.0017214485641282$ &$  0.0069288505780870$ &$  0.0032237126853854$ \\
$ 13$ &$   0.0011215139675216$ &$  0.0040732105791174$ &$  0.0022173172757962$ \\
$ 14$ &$   0.0001704882398608$ &$ -0.0000794545807413$ &$  0.0016753488483300$ \\
$ 15$ &$  -0.0002998652087043$ &$ -0.0007340446903331$ &$ -0.0011606884628785$ \\
$ 16$ &$  -0.0001452851719807$ &$ -0.0004547873303444$ &$ -0.0004060678489517$ \\
$ 17$ &$  -0.0000920708232117$ &$ -0.0002478140795869$ &$ -0.0000899247570665$ \\
$ 18$ &$   0.0000577763210731$ &$  0.0001850306275303$ &$  0.0000772987554511$ \\
$ 19$ &$   0.0000179810577610$ &$  0.0000655287922988$ &$  0.0000297523799899$ \\
$ 20$ &$   0.0000073202860851$ &$  0.0000327013420245$ &$  0.0000079670215055$ \\
$ 21$ &$  -0.0000041801789562$ &$ -0.0000194432981798$ &$ -0.0000053190023630$ \\
$ 22$ &$  -0.0000011935254922$ &$ -0.0000057711652336$ &$ -0.0000017383936002$ \\
$ 23$ &  $-$\qquad\qquad\qquad  &$ -0.0000010832435911$ &$ -0.0000010832435911$ \\
$ 24$ &  $-$\qquad\qquad\qquad  &$  0.0000006185758332$ &$  0.0000006185758332$ \\
$ 25$ &  $-$\qquad\qquad\qquad  &$  0.0000001766158898$ &$  0.0000001766158898$ \\
	\hline
    \end{tabular}
    \caption{Coefficients as above for $W_{872}$.
    }
    \label{table:c872}
\end{table*}
\begin{table*}[]
    \begin{tabular}{|c||r|r|r|}
	\hline
	$i$ & $c_i$ \qquad \qquad \qquad & mid \quad \quad \quad & high\qquad \qquad \qquad \\ 
	\hline
$ 0$ &$   0.6339678630846935$ &$ -0.2712444288646899$ &$ -0.4376120240714272$ \\
$ 1$ &$   0.4937325799533404$ &$  0.4039182339362101$ &$  0.4425723917378794$ \\
$ 2$ &$   0.1947092720525292$ &$  0.4039182339362101$ &$ -0.4425723917378794$ \\
$ 3$ &$  -0.0475399624330072$ &$ -0.2712444288646899$ &$  0.4376120240714272$ \\
$ 4$ &$  -0.1058504106212748$ &$ -0.4003612905337904$ &$ -0.1482339583941647$ \\
$ 5$ &$  -0.0388674200543010$ &$  0.2836382952046013$ &$ -0.2252241642218347$ \\
$ 6$ &$   0.0280045599991956$ &$ -0.0869402103053695$ &$  0.1415067231884696$ \\
$ 7$ &$   0.0360267579988863$ &$  0.0808627388944033$ &$  0.1146487108662731$ \\
$ 8$ &$   0.0064755478963359$ &$  0.0625113141364469$ &$ -0.0779159680299839$ \\
$ 9$ &$  -0.0114606959238103$ &$ -0.0542988486674284$ &$  0.0065857070698602$ \\
$ 10$ &$  -0.0105211360558507$ &$ -0.0375578779290256$ &$ -0.0198322246382788$ \\
$ 11$ &$  -0.0004029241837836$ &$  0.0067718607880278$ &$ -0.0080093797622695$ \\
$ 12$ &$   0.0033019830277673$ &$  0.0072289136257594$ &$  0.0116541898416100$ \\
$ 13$ &$   0.0023926174915384$ &$  0.0072590112195473$ &$  0.0067065096826978$ \\
$ 14$ &$   0.0001341991885869$ &$  0.0027315433528202$ &$ -0.0002487510217457$ \\
$ 15$ &$  -0.0007538488057219$ &$ -0.0033602186014616$ &$ -0.0017279646580792$ \\
$ 16$ &$  -0.0004329594781529$ &$ -0.0016600052145813$ &$ -0.0010614642726768$ \\
$ 17$ &$  -0.0001478738121134$ &$ -0.0002029246011991$ &$ -0.0007777985405009$ \\
$ 18$ &$   0.0001677730970039$ &$  0.0004669843226612$ &$  0.0005667275560345$ \\
$ 19$ &$   0.0000705400971549$ &$  0.0002308672126553$ &$  0.0002018690639151$ \\
$ 20$ &$   0.0000462924730060$ &$  0.0001574234886506$ &$  0.0000672760624394$ \\
$ 21$ &$  -0.0000318321017008$ &$ -0.0001118649601750$ &$ -0.0000576398269189$ \\
$ 22$ &$  -0.0000102706557188$ &$ -0.0000377747073410$ &$ -0.0000208603073810$ \\
$ 23$ &$  -0.0000059656665643$ &$ -0.0000191233742666$ &$ -0.0000154881935497$ \\
$ 24$ &$   0.0000034648630245$ &$  0.0000126260685159$ &$  0.0000087886962940$ \\
$ 25$ &$   0.0000009680025867$ &$  0.0000038839469583$ &$  0.0000024869174397$ \\
$ 26$ &$   0.0000005103695558$ &$  0.0000025108247769$ &$  0.0000008948821449$ \\
$ 27$ &$  -0.0000002386702849$ &$ -0.0000012807717653$ &$ -0.0000005250889782$ \\
$ 28$ &$  -0.0000000558061352$ &$ -0.0000003243977491$ &$ -0.0000001477032107$ \\
$ 29$ &  $-$\qquad\qquad\qquad  &$ -0.0000000805960182$ &$ -0.0000000805960182$ \\
$ 30$ &  $-$\qquad\qquad\qquad  &$  0.0000000376900903$ &$  0.0000000376900903$ \\
$ 31$ &  $-$\qquad\qquad\qquad  &$  0.0000000088127363$ &$  0.0000000088127363$ \\
 
	\hline
    \end{tabular}
    \caption{Coefficients as above for $W_{1092}$.
    }
    \label{table:c1092}
\end{table*}

\end{document}